\documentclass[11pt,twocolumn]{article}
\usepackage{authblk}
\usepackage{fullpage}
\usepackage{amssymb,amsmath}
\usepackage{soul}
\usepackage[T1]{fontenc}
\usepackage{siunitx}
\usepackage[version=3]{mhchem}
\usepackage{xr}
\usepackage{bm,bbm}
\usepackage{pdfpages}
\usepackage{float}
\usepackage{graphicx}
\usepackage{dcolumn}
\usepackage{bm}
\usepackage{color}
\usepackage{tikz}
\usepackage{setspace}
\usepackage{cancel}
\usepackage[title]{appendix}
\usepackage[font=small,center,labelfont=bf]{caption}
\usepackage{float}
\usepackage{braket}

\usepackage[lof,lot]{etoc}

\usepackage[unicode=true]{hyperref}
\hypersetup{colorlinks=true,
            citecolor=blue,
            linkcolor=blue}
            
\usepackage[square,numbers,sort&compress]{natbib}
\usepackage{setspace}

\numberwithin{equation}{section}

\usepackage[switch]{lineno}
\usepackage[left=1.5cm,right=1.5cm,top=3.0cm,bottom=3.0cm]{geometry}
\newcommand*\linenomathpatch[1]{%
  \cspreto{#1}{\linenomath}%
  \cspreto{#1*}{\linenomath}%
  \csappto{end#1}{\endlinenomath}%
  \csappto{end#1*}{\endlinenomath}%
}

\linenomathpatch{equation}
\linenomathpatch{gather}
\linenomathpatch{multline}
\linenomathpatch{align}
\linenomathpatch{alignat}
\linenomathpatch{flalign}

\floatname{algorithm}{Algoritmo}

\makeatletter

\newcommand{\hypgeo}[2]{%
  {\vphantom{F}}_{#1}\kern-\scriptspace F_{#2}%
}

\newcommand\approxsim{\mathchoice
  {\@approxsim {\displaystyle}      {1ex} }
  {\@approxsim {\textstyle}         {1ex} }
  {\@approxsim {\scriptstyle}       {.7ex}}
  {\@approxsim {\scriptscriptstyle} {.5ex}}}
\newcommand\@approxsim[2]{%
  \mathrel{%
    \ooalign{%
      $\m@th#1\sim$\cr
      \hidewidth$\m@th#1.$\hidewidth\cr
      \hidewidth\raise #2 \hbox{$\m@th#1.$}\hidewidth\cr
    }%
  }%
}

\newcommand{\bo}{\raise-1mm\hbox{\Large$\Box$}}

\newcommand{\ed}[1]{{\color{black}{#1}}}

\title{How spatial patterns can lead to less resilient ecosystems}

\author[1]{David Pinto-Ramos}
\author[1,2*]{Ricardo Martinez-Garcia}

\affil[1]{\small \it Center for Advanced Systems Understanding (CASUS) -- Helmholtz-Zentrum Dresden-Rossendorf (HZDR), Germany}
\affil[2]{\it ICTP South American Institute for Fundamental Research \& Instituto de F\'isica Te\'orica, Universidade Estadual Paulista - UNESP, Brazil.}
\affil[*]{Corresponding author: r.martinez-garcia@hzdr.de}
\date{}

\begin{document}
\twocolumn[
\begin{@twocolumnfalse}
\maketitle	


\begin{abstract}
Several theoretical models predict that spatial patterning increases ecosystem resilience. However, these predictions rely on simplifying assumptions, such as assuming isotropic and infinitely large ecosystems, and empirical \ed{evidence directly linking spatial patterning to enhanced resilience remains scarce.} We introduce a unifying framework, encompassing existing models for vegetation pattern formation in water-stressed ecosystems, that relaxes these assumptions. This framework incorporates finite vegetated areas surrounded by desert and anisotropic environmental conditions that lead to non-reciprocal plant interactions. Under these more realistic conditions, we identify a novel desertification mechanism, known as \ed{nonlinear} convective instability in physics but largely overlooked in ecology. These instabilities form when non-reciprocal interactions destabilize the vegetation–desert interface and can trigger desertification fronts even under stress levels where isotropic models predict stability. Importantly, ecosystems exhibiting periodic vegetation patterns are more susceptible to \ed{nonlinear} convective instabilities than those with homogeneous vegetation, suggesting that spatial patterning may reduce, rather than enhance, resilience. \ed{These findings challenge the prevailing view that self-organized patterning enhances ecosystem resilience and provide a new framework for investigating how spatial dynamics shape the stability and resilience of ecological systems under changing environmental conditions.}
\end{abstract}

\newline
 \end{@twocolumnfalse}
]
\normalsize

\clearpage

\section{Introduction}
{E}cosystems can exhibit multiple alternative stable states, each with different structures and functions \cite{Beisner2003,Schroder2005}. Due to the nonlinear processes underlying ecosystem dynamics, small changes in environmental conditions can trigger persistent changes between these states once a critical threshold, or tipping point, is crossed \cite{Scheffer2009a}. These transitions, known as regime shifts, may propagate over large distances and even different biomes \cite{Rocha2018}, impacting ecosystem services and human well-being substantially \cite{Biggs2012, Kraberg2011}. Given the severe consequences of regime shifts and the challenges associated with predicting or reversing them, much research has focused on developing predictive theories of regime shifts as well as quantitative indicators to anticipate and potentially mitigate their effects \cite{Martin2015,Kefi2014,Scheffer2009}.

Regime shifts have been observed in different terrestrial and marine ecosystems, including shallow lakes, savannas, kelp forests, or drylands \cite{Folke2004}. In drylands, regime shifts occur when aridity crosses a critical threshold, causing vegetation loss and ecosystem collapse into a desert state \cite{berdugo2020}. Models suggest that spatial processes\textemdash particularly self-organized regular patterns of vegetation and bare soil\textemdash shape dryland adaptive capacity, enhancing ecosystem resilience and robustness against external perturbations \cite{VandeKoppel2004,rietkerk2021evasion}. 

First, these patterns emerge once a resource-scarcity threshold is crossed and adjust their shape as resource availability further decreases \cite{Deblauwe2011,VonHardenberg2001}, suggesting that patterns could be a signature of stress and indicate an ongoing desertification process and proximity to a full vegetation collapse \cite{rietkerk2004self}. Second, ecosystems exhibiting periodic patterns can persist beyond the tipping point predicted by non-spatial theories, thereby enhancing their resistance at higher aridity levels \cite{Siteur2014}. Finally, patterns with slightly different spatial properties can coexist across a range of environmental conditions, suggesting that patterning is an adaptive feature that allows ecosystems to buffer external perturbations \cite{rietkerk2021evasion,bastiaansen2018multistability,Bastiaansen2020}. 

However, \ed{recent modeling work in savannas questions this view, suggesting that, for very specific parameter combinations, self-organized patterns may represent transient and unstable states that facilitate, rather than prevent, ecosystem collapse \cite{van2025vegetation}.} More importantly, because pattern formation and vegetation dynamics occur over large spatial and temporal scales, opportunities to empirically test all these theoretical predictions remain limited \cite{bastiaansen2018multistability,Veldhuis2022}. Consequently, whether and how spatial patterning influences ecosystem resilience remains unknown \cite{Rietkerk2021,Tarnita2024}.

The relationship between vegetation patterns and desertification processes has been mainly studied using simple models for infinite vegetated areas in flat, isotropic landscapes. However, real vegetation patterns cover large, but finite, regions \cite{clerc2021localised}, and are often embedded in landscapes with different topographies, such as hillsides or microreliefs \cite{Wilson1964, Florinsky1996, Valentin1999,gandhi2018topographic}. \ed{Although spatially isotropic models often reproduce observed vegetation patterns \cite{Deblauwe2011}, topography and other environmental features, such as wind or directional fog, can introduce spatial anisotropies. These anisotropies make interactions between vegetation patches depend not only on distance but also on relative position, turning them non-reciprocal. More importantly, assuming infinite landscapes neglects several spatial processes that can cause vegetation loss and tipping, the most important being the propagation or reversal of desertification fronts in response to increasing aridity \cite{zelnik2017desertification,fernandez2019front}. }

Desertification front propagation can occur before tipping points, turning abrupt regime shifts into gradual \cite{bel2012gradual}, and is sensitive to boundaries, spatial heterogeneities, and the size of the vegetated area. Spatial heterogeneity in plant interactions and boundary effects have been introduced separately in studies of vegetation pattern formation to explain the formation of vegetation stripes \cite{lefever1997origin,klausmeier1999regular} or the spreading of invasive species \cite{bennett2023evidence}, among other phenomena. Yet, only recently have studies included both to successfully explain topological properties observed in banded vegetation patterns worldwide \cite{pinto2023topological}. How much they influence dryland stability remains to be explored.

We address this gap by investigating the effects of spatial heterogeneity and non-reciprocity, induced by boundary effects and environmental anisotropies, on desertification dynamics. Our results show that non-reciprocity in plant interactions enhances the invasion of desertification fronts into vegetated areas, thereby increasing the likelihood of regime shifts at lower environmental stress. Moreover, this phenomenon is more pronounced in ecosystems with vegetation patterns, causing self-organized systems to collapse under stress levels that non-patterned vegetation can survive. This result suggests that spatial self-organization may make ecosystems more susceptible to regime shifts and less resistant to environmental change, challenging the current consensus that spatial patterning enhances system resilience.

\section{A unifying model for vegetation patterns}\label{sec:framework}

\noindent \textbf{Model equation.} Several models have been proposed to describe vegetation dynamics in water-limited ecosystems. We will focus on models describing vegetation as a continuous biomass density $b(x,t)$ evolving in time and space according to partial differential equations (PDE), which is the usual choice to study regular vegetation patterns \cite{meron_continuum_2019}. 

\begin{figure*}[t!]
	\centering
	\includegraphics[width=0.7\textwidth]{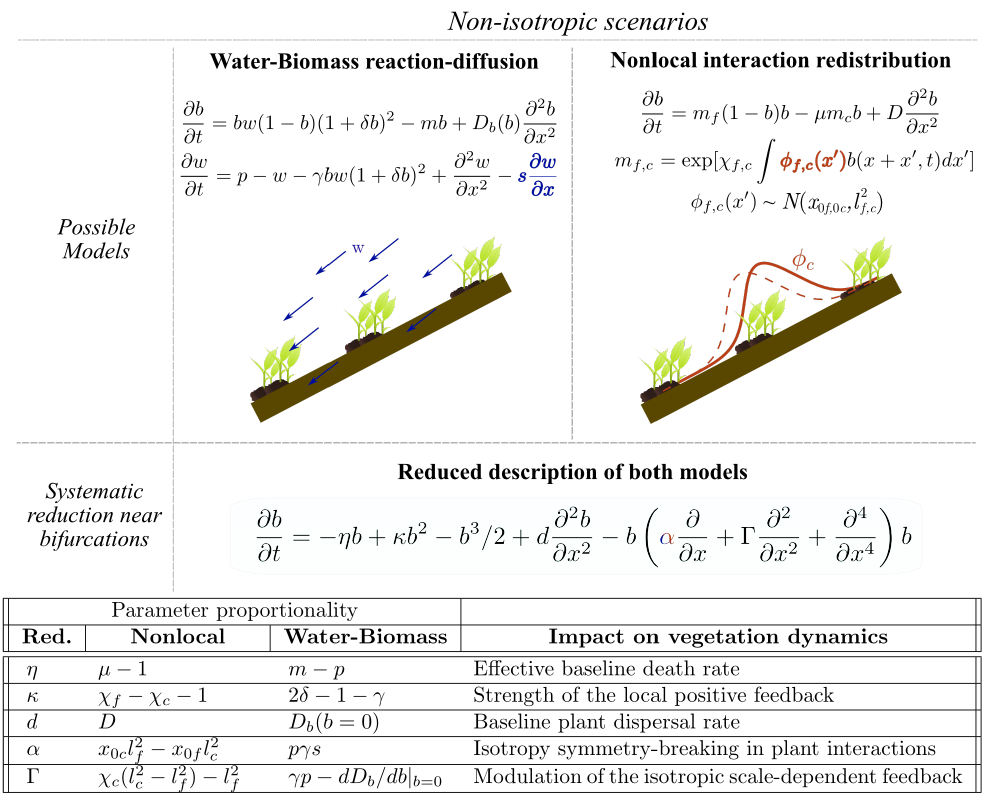}
	\caption{Schematic summary of model reduction from two representative examples of reaction-diffusion and nonlocal interaction-redistribution models to the reduced equation \eqref{eq1}. The parameters of the reduced equation encapsulate the ecological feedbacks in each of the original models as described in the table.}
	\label{F1}
\end{figure*}

The diversity of PDE-based models for vegetation dynamics can be organized into two main categories: nonlocal interaction redistribution models, in which plant interactions are implicitly described by kernel functions that modulate the intensity of positive and negative density-dependence in plant growth, and reaction-diffusion models that explicitly describe water-vegetation feedbacks \cite{MartinezGarcia2023}. Although they are mathematically different, models within these two categories can be reduced to a general equation near the onset of instability of the unpopulated state (Fig.\,\ref{F1}). 

We exploit this \textit{universality} and perform our analyses using the reduced equation, which makes our results independent of specific modeling assumptions. We provide a full derivation of this equation in Supporting Information App. A, starting from two prototypical examples of kernel-based and reaction-diffusion models, while we focus here on presenting the reduced model and discussing its general structure. In its dimensionless form, the reduced equation reads
\begin{eqnarray}
\frac{\partial b}{\partial t} &=& -\eta b + \kappa b^2 - \frac{b^3}{2} + d\frac{\partial^2 b}{\partial x^2} \nonumber\\
&&-b\left(\alpha \frac{\partial}{\partial x} +\Gamma \frac{\partial^2}{\partial x^2}+\frac{\partial^4}{\partial x^4}\right)b,
\label{eq1}
\end{eqnarray}
where the specific ecological meaning of each parameter will depend on which model we use to derive it (see Supporting Information App.\,A and Fig.\,\ref{F1}). \ed{Assuming additional constraints, an equation similar to Eq.\,\eqref{eq1} was derived in Ref.\,\cite{pinto2023topological} for a two-dimensional system. Although the calculations in \cite{pinto2023topological} require additional constraints compared to ours and were performed only for the non-local model, their equation reduces to \eqref{eq1} in the one-dimensional limit.} 

In general, Eq.\,\eqref{eq1} consists of a linear net death term  $-\eta b$, where $\eta$ can be negative to represent growth, and local competition and facilitation contained in the terms $\kappa b^2-b^3/2$. Ecologically, these first three terms define a nonlinear density-dependent growth rate, for example, due to an Allee effect. The spatial dynamics is described by a linear diffusion accounting for plant dispersal with diffusion constant $d$, and a nonlinear spatial differential operator. In this nonlinear operator, the term proportional to $\alpha$ originates from isotropy-breaking processes, while the last two terms encapsulate the pattern-forming feedbacks. Specifically, the combination of the linear and density-dependent nonlinear diffusion leads to an effective diffusion term, $d-b\Gamma$, that can become negative and trigger spontaneous aggregation of the biomass for sufficiently large $\Gamma$. This aggregation process is saturated by the last term, resulting in stable patterns. 

Three parameters in Eq.\,\eqref{eq1} are particularly relevant for our analysis. First, the net death rate $\eta$ measures the difference between baseline death and growth rates and hence serves as a proxy for environmental stress. Large positive values of $\eta$ represent high stress, and stress decreases as $\eta$ decreases, eventually becoming negative when environmental conditions are such that they can sustain baseline population growth. Second, $\alpha$ modulates the only term breaking the isotropy in Eq.\,\eqref{eq1} and hence controls the intensity of nonreciprocity in plant-plant interactions. This symmetry-breaking can be caused by various factors, such as slopes, more complex but uni-directional topographies, or fog water carried by wind \cite{klausmeier1999regular,gandhi2018topographic,hidalgo2024nonreciprocal}. Lastly, $\Gamma$ directly controls the pattern formation instability, allowing us to study homogeneous or patterned ecosystems by varying this single parameter.\\ 

\noindent \textbf{Boundary conditions.} To fully describe the model, we need to specify the boundary conditions for Eq.\,\eqref{eq1}. Most studies assume periodic boundary conditions, mimicking infinitely large and spatially invariant ecosystems. This assumption, while mathematically convenient, is not realistic in most ecosystems, where vegetation (patterned or not) is often embedded in non-vegetated areas or at the edge of a dryland desert transition \cite{Deblauwe2011,clerc2021localised}. To model these scenarios, we use Dirichlet boundary conditions, which set vegetation biomass to zero at the system edges, $x\leq0$ and $x\geq L$ (with $L$ the system length). \ed{These boundary conditions assume that the environmental stress $\eta$ grows sharply to infinity at the ecosystem boundaries.}

Because of the biomass-vanishing boundary conditions, any solution in the ecosystem bulk must connect with the unvegetated state ($b=0$) at the boundaries, which is also a solution of Eq.\,\eqref{eq1}. Consequently, a front necessarily forms for all the solutions except the unvegetated one, which is the only strictly homogeneous solution possible in the system. The motion of this front will dictate whether the desert state $b=0$ propagates into the vegetated region or not. This motion is directly determined by the parameters of Eq.\,\eqref{eq1}, which account for the strength of the different plant interactions and environmental stresses. We next analyze this front-propagation process, both in cases where plant biomass in the vegetated area is quasi-homogeneous (i.e., homogeneous within the bulk) and in cases where it forms self-organized regular patterns.

\section{Results}

\subsection{Stationary solutions \ed{and pattern-formation instability}}\label{sec:homogeneous}

Because boundary effects smoothly take vegetation density to zero at the edges of the vegetated patch, our model does not allow for strictly homogeneous states with $b\neq 0$. However, far from these vegetation-desert boundaries (i.e., in the bulk of the vegetated area), we can neglect boundary effects and obtain quasi-homogeneous solutions from the nonspatial terms of Eq.\,\eqref{eq1}. These solutions are $b_{\pm}(\eta, \kappa) = \kappa \pm \sqrt{ \kappa^2-2\eta}$ and $b_0=0$, where $\kappa$ controls the existence of alternative stable states by modulating the intensity of facilitation, and the linear mortality rate $\eta$ is a proxy for the intensity of environmental stress. 

For $\kappa<0$ (i.e., no net facilitation), the nonspatial model is monostable. If, additionally, $\eta<0$ (low stress), then $b_0$ is unstable, and $b_+$ is stable. At higher stress $\eta>0$, $b_0$ is the only possible steady state and thus stable. In the presence of facilitation, $\kappa>0$, the system presents two alternative stable states for stress levels $0<\eta<\eta_{\mathrm{s.n}}$, with $\eta_{\mathrm{s.n}} = \kappa^2/2$ (black curves in Fig.\,\ref{F2}a,\,b). $\eta_{\mathrm{s.n}}$ thus represents a tipping point: when environmental conditions deteriorate beyond this value, the system undergoes a sudden regime shift from a vegetated state $b_+$ to a desert $b_0$. Near the edges of the bulk, the homogeneous solution $b_+$ connects to the boundary condition $b=0$, creating two interfaces, or fronts, one near $x=0$ and the other near $x=L$. Because $\eta$ sets the level of environmental stress by setting the effective plant mortality of the plants, it acts as the control parameter. 

Patterned solutions are much harder to obtain analytically, but we can determine the region of parameter space where they occur by performing a linear stability analysis of Eq.\,\eqref{eq1}. Again, our analysis focuses on the ecosystem bulk and therefore assumes that the boundaries are infinitely far apart. To perform the linear stability analysis, we introduce a small perturbation to the uniform solution, $b=b_+ + u_0e^{\lambda t + ikx}$ and linearize in $u_0$, which gives a perturbation growth rate of the form
\begin{equation}
    \lambda(k) = -\eta +2\kappa b_+ - 3b_+^2/2 -(d-b_+\Gamma)k^2 - b_+k^4 -i b_+ \alpha k.
\end{equation}
A heterogeneous solution may appear whenever $\lambda(k)>0$ for some $k\neq 0$, meaning that the term $\propto e^{ikx}$ will start to grow exponentially. This condition is true when the environmental stress is higher than a threshold value $\eta_\mathrm{T}$, which we can obtain by simultaneously solving
\begin{eqnarray}
    k_c^2&=& (b_\mathrm{T}\Gamma-d)/ (2b_\mathrm{T}),\nonumber \\
    0&=&-\eta_\mathrm{T} +2\kappa b_\mathrm{T} - 3b_\mathrm{T}^2/2+b_\mathrm{T}k_c^4,
\end{eqnarray}
with $b_\mathrm{T}=b_+(\eta_\mathrm{T}, \kappa)$. This analysis allows us to identify $\Gamma$ and $d$ as the parameters triggering a Turing instability and thus patterns ($k_c^2$ must be positive). In the following sections, we numerically investigate the stability of patterned and quasi-homogeneous vegetation distributions in response to worsening environmental conditions (see \nameref{sec:MatMet} for a description of the numerical method). We perform these analyses using the reduced Eq.\,\eqref{eq1}, but our results also hold for the original kernel-based and reaction-diffusion models (see Supporting Information App.\,\ed{B and Fig.\,S1.}).

\subsection{Desertification dynamics}

To analyze the persistence or tipping of vegetated states under worsening environmental conditions, \ed{we consider an ecosystem under low environmental stress $\eta=-0.05$ and trace its fate as $\eta$ increases quasi-statically, i.e., slowly enough to allow the system to reach equilibrium between consecutive changes in $\eta$ (see \nameref{sec:MatMet} for details). 

We analyze three increasingly complex scenarios. First, we recover in our model well-known results for infinite (or periodic) ecosystems with reciprocal plant interactions. Next, we introduce the effect of surrounding deserts by replacing periodic with Dirichlet boundary conditions while maintaining reciprocity in plant interactions. Finally, we address the full problem, incorporating both finite vegetated areas and nonreciprocal interactions ($\alpha\neq0$). In each case, we compare scenarios without spatial patterning (small $\Gamma$) to those where patterns can form (large $\Gamma$).}

\subsubsection{Desertification by vegetation destabilization: \ed{uniform ecosystem tipping}.}
\begin{figure}[t!]
	\centering
	\includegraphics[width=0.9\columnwidth]{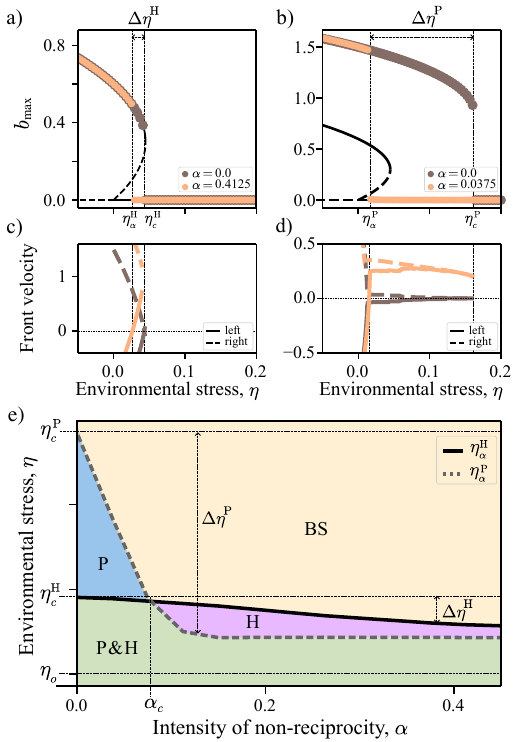}
	\caption{a,\,b) \ed{Bifurcation diagram and front velocity for quasi-homogeneous vegetation ($\Gamma=1$) under varying environmental stress $\eta$ (see \nameref{sec:MatMet} for details). Brown (dark) and orange (light) symbols/curves denote reciprocal and non-reciprocal plant interactions, respectively. c,\,d) Same as a,b, but for patterned vegetation ($\Gamma=2.25$). In all panels, $\eta_{(c,\alpha)}^{\text{\tiny{(H,P)}}}$ mark tipping points for quasi-homogeneous (H) and patterned (P) vegetation with reciprocal ($c$) and non-reciprocal ($\alpha$) interactions. e) In the $(\eta,\alpha)$ space, curves $\eta_{\alpha}^{\tiny\text{\tiny{(H,P)}}}$\textemdash dashed for patterned and solid for non-patterned vegetation\textemdash delineate regime boundaries where different spatial configurations persist (P: patterns, H: quasi-homeogeneous, BS: bare soil). $\alpha_c$ denotes the non-reciprocity threshold beyond which spatial patterning reduces resilience.}} 
	\label{F2}
\end{figure}

\ed{We first show how our model recovers well-known results in the classical scenario with periodic boundary conditions and reciprocal plant interactions, $\alpha = 0$. When $\Gamma$ is small and patterns cannot form, the system behaves as described in Section \ref{sec:homogeneous}. Therefore, as environmental stress increases gradually, vegetation biomass declines uniformly following the $b_+$ solution (solid black curve in Fig.\,\ref{F2}a,\,b) and collapses abruptly once $\eta$ exceeds $\eta_{\mathrm{s.n}}$.

\ed{At large $\Gamma$ the system behaves differently. Past an environmental stress threshold $\eta_\mathrm{T}<\eta_{\mathrm{s.n}}$, the system crosses a Turing instability and local perturbations can trigger spatial patterns. This branch of patterned configurations enhances vegetation biomass relative to the homogeneous case and collapses abruptly when environmental stress exceeds a threshold that is typically higher than $\eta_{\mathrm{s.n}}$ \cite{VonHardenberg2001,zelnik2017desertification,Moreno2025}. The difference between the tipping points of patterned and non-patterned vegetation is interpreted as evidence that spatial patterns increase ecosystem resilience \cite{Rietkerk2021}.}

\subsubsection{Desertification by spatiotemporal structures in the reciprocal regime}

Next, we study the effect of introducing finite vegetated areas while keeping reciprocal plant interactions. Mathematically, this is achieved with the Dirichlet boundary conditions introduced in Section \ref{sec:framework}, which assume an infinite environmental stress, and consequently $\eta\rightarrow\infty$, at the ecosystem borders. 

In this scenario, if the environmental stress within the ecosystem is sufficiently low, an initial patch of vegetation grows and propagates until it reaches the system edges ($x=\lbrace 0, L\rbrace$) and is stopped by the infinite mortality. Therefore, boundary conditions create two fronts connecting vegetated and bare-soil states, and although their velocities are such that the vegetated state would continue to propagate, this expansion is blocked at equilibrium by the boundary conditions. In response to increasing environmental stress, these fronts can flip their velocity and propagate bare-soil over the vegetated state, triggering a desertification process \cite{zelnik2017desertification}. 

When $\Gamma$ is small enough to prevent spatial patterning for any $\eta$, the fronts remain stable and the vegetation biomass within the ecosystem bulk follows $b_+$ as environmental stress increases (brown symbols in  Fig.\,\ref{F3}a). However, upon crossing a threshold in environmental stress, front velocities change sign, triggering the propagation of a desertification front that causes ecosystem collapse (brown curves in Fig.\,\ref{F2}c; see \nameref{sec:MatMet} for details on how we computed front velocities). This critical parameter value where front velocities vanish, $\eta_m^\text{\tiny{H}}$, is usually called the Maxwell point \cite{pismen2006patterns,zelnik2018regime}. 
Hence, non-patterned ecosystems in this limit collapse due to the reversal of the vegetation-desert front velocity, and the point of ecosystem tipping, $\eta_c^\text{\tiny{H}}$, satisfies $\eta_c^\text{\tiny{H}} = \eta_m^\text{\tiny{H}}<\eta_\text{s.n}$.

For larger $\Gamma$, increasing environmental stress eventually triggers spatial patterns (the Turing point $\eta_\mathrm{T}$ is not shown in Fig.\,\ref{F2}b because $\eta_\mathrm{T}<0.05$ for $\Gamma = 2.25$). Consequently, vegetation biomass within the bulk increases relative to the non-patterned case, $b_\mathrm{\tiny{max}}>b_+$ (brown symbols in Fig.\,\ref{F3}b), until it collapses upon reaching a stress level $\eta_c^\text{\tiny{P}}>\eta_c^\text{\tiny{H}}$. In this case, the velocity of fronts connecting patterned vegetation and bare soil vanishes at a stress value, $\eta_m^\text{\tiny{P}}$, much lower than the threshold for ecosystem collapse, $\eta_m^\text{\tiny{P}}<\eta_c^\text{\tiny{P}}$. Moreover, these velocities never change sign (brown curves in Fig.\,\ref{F3}d). This result shows that between $\eta_m^\text{\tiny{P}}$\textemdash which now indicates the boundary of the pinning region\textemdash and $\eta_c^\text{\tiny{P}}$, the pattern consists of several localized structures. At $\eta_c^\text{\tiny{P}}>\eta_\text{s.n}>\eta_c^\text{\tiny{H}}$, the ecosystem collapses due to uniform tipping rather than front propagation for these parameter values.

Therefore, in this scenario with finite vegetated areas and reciprocal plant interactions, spatial patterns still confer enhanced resilience to vegetated areas and make them resistant to invading desertification fronts \cite{VonHardenberg2001,lejeune2002localized,Siteur2014,rietkerk2021evasion}. Notice, however, that we obtained this result by varying quasi-statically the environmental stress control parameter, such that the solution branches in Fig.\,\ref{F3}a,\,b (brown dots) represent the observed equilibria of the system. More complex ways of varying the control parameter, such as parameter quenches or temporal variations, as well as intricate initial conditions, would lead to a much more complex diagram of states \cite{zelnik2013regime,zelnik2018regime,bel2012gradual,bordeu2016self,tlidi2018extended,echeverria2023effect,clerc2021localised}.

}

\subsubsection{Desertification by \ed{nonlinear convective instabilities induced by non-reciprocity}.}

\ed{We finally study the full problem, introducing finite vegetated areas and non-reciprocal plant interactions}, $\alpha\neq 0$. We consider only $\alpha>0$, which is equivalent to $\alpha<0$ under a $x\rightarrow -x$ reflection of the spatial coordinate. Regardless of whether vegetation is quasi-homogeneous or forms spatial patterns, non-reciprocal interactions anticipate the ecosystem collapse to a lower stress threshold $\eta_{\alpha}<\eta_{c}$ (orange vs. brown symbols in Fig.\,\ref{F2}a,\,b) \ed{and thus reduce ecosystem resistance to worsening environmental conditions, $\Delta\eta = \eta_{c}-\eta_{\alpha}$. 

The advection term in Eq.\,\eqref{eq1}, responsible for non-reciprocities, introduces a new component into both front velocities. This component goes to the right because $\alpha>0$ and, once the environmental stress crosses a threshold value $\eta_{\alpha}<\eta_c$, makes both fronts move in the same direction (see how both orange curves are shifted to more positive values in Fig.\,\ref{F2}d). As a result, a desertification front propagates from left to right, causing ecosystem collapse. This inversion in the front velocity is known as a nonlinear convective instability \cite{chomaz1992absolute}, and has been reported in other nonlinear systems driven by non-reciprocities\cite{pinto2025exact,veenstra2024nonreciprocal,aguilera2024nonlinear} (see \nameref{sec:MatMet} for an analytical argument of how non-reciprocities can trigger this instability). Patterned systems are more sensitive to this phenomenon because the velocity of the fronts connecting patterned vegetation to desert at $\alpha = 0$ is much lower than that of the fronts connecting homogeneous vegetation to desert at the same stress level(Fig.\,\ref{F2}c,\,d)}. 


To better understand how non-reciprocal plant interactions destabilize vegetated states, we \ed{computed the environmental stress triggering ecosystem collapse for various values of the non-reciprocity parameter and both homogeneous and patterned configurations ($\eta_\alpha^\text{\tiny{H}}$ solid and $\eta_\alpha^{\text{\tiny{P}}}$ dashed curves in} Fig.\,\ref{F2}e, respectively). Although increasing non-reciprocity in plant interactions leads to earlier ecosystem collapses regardless of vegetation spatial distribution, patterned ecosystems are more sensitive to increasing $\alpha$ than quasi-homogeneous ones. For weak non-reciprocity, patterned ecosystems are more resistant than non-patterned ones $\eta_\alpha^{\text{\tiny{P}}}>\eta_\alpha^\text{\tiny{H}}$, consistent with existing theory \cite{Rietkerk2021}. However, because $\eta_{\alpha}^{\text{\tiny{P}}}$ decays with increasing $\alpha$ much faster than $\eta_{\alpha}^{\text{\tiny{H}}}$, there is a crossover between both curves at a critical strength of the non-reciprocity parameter $\alpha_c$. Beyond $\alpha_c$, patterned ecosystems collapse under lower environmental stress than those with quasi-homogenous vegetation ($\eta_\alpha^\text{\tiny{H}}>\eta_\alpha^{\text{\tiny{P}}}$). 
This result suggests that spatial self-organization could hinder ecosystem resistance to worsening environmental conditions, which challenges the prevailing consensus that it enhances robustness and resilience.

\ed{This result still holds in the original models used to derive the reduced equation (see Supporting Information App. B), as well as in two-dimensional systems and when $\eta$ increases smoothly at the ecosystem boundaries (see Supporting Information App.\,C and D). Additionally, using parameter estimates available in the literature \cite{meron2015ModelNew,cain1997clonal}, we parameterized the original reaction-diffusion water-vegetation model and found that spatial patterns can lead to less resistant ecosystems for intensities of non-reciprocity that are compatible with empirical estimates of water advection intensity reported in previous studies (see Supporting Information App. E; \cite{klausmeier1999regular})}

\begin{figure}[t!]
	\centering
	\includegraphics[width=0.46\textwidth]{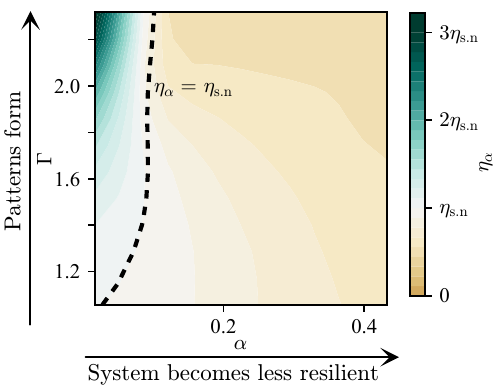}
	\caption{Behavior of the tipping point \ed{in the non-reciprocal system $\eta_{\alpha}$ as a function of $\Gamma$ and $\alpha$. Increasing $\Gamma$ triggers the Turing instability and changes the pattern wavelength and amplitude, while increasing $\alpha$ always destabilizes the ecosystem. The color map measures $\eta_{\alpha}$ relative to $\eta_\mathrm{{s.n}}$, and a dashed line indicates the limit $\eta_{\alpha}=\eta_\mathrm{{s.n}}$, separating regions where spatial effects make vegetation more or less resilient compared to the result of non-spatial models.}}
	\label{F3}
\end{figure}

\ed{Finally, we quantified how much and under what conditions pattern features and non-reciprocal interactions jointly impact the environmental stress threshold triggering ecosystem collapse. We computed the convective instability threshold, $\eta_\alpha$, as a function of $\Gamma$\textemdash main responsible for pattern wavelength and amplitude\textemdash and $\alpha$, and compared these values with the tipping point of the system when spatial effects are neglected, $\eta_{\text{s.n}}$. For weakly non-reciprocal interactions, $\eta_{\alpha}>\eta_{\text{s.n}}$}, indicating that self-organized patterns enhance ecosystem resistance (teal, top left region of Fig.\,\ref{F3}). However, past a limit value of the non-reciprocity parameter, \ed{ $\eta_{\alpha}<\eta_{\text{s.n}}$}, indicating that spatial effects reduce ecosystem resistance. \ed{The curve separating these two regions, $\eta_\alpha=\eta_\text{s.n}$, tends to an asymptotic value of $\alpha$ as $\Gamma$ increases (black dashed line in Fig.\,\ref{F3}). This asymptote defines an upper bound on the intensity of non-reciprocal interactions such that ecosystems with non-reciprocity exceeding this value are less resilient than their non-spatial counterparts, regardless of $\Gamma$. Moreover, ecosystems become more sensitive to $\alpha$ deeper into the Turing instability region, because $\eta_{\alpha}$ takes greater values when $\alpha\rightarrow 0$, and they rapidly decrease as $\alpha$ is increased (the contour levels are tightly packed, indicating high gradients).}

\section{Discussion}

Using a reduced equation representative of two mainstream models for vegetation pattern formation, we analyzed spatiotemporal vegetation dynamics in confined environments where environmental anisotropies lead to non-reciprocal plant interactions. These two factors\textemdash spatial confinement and directional environmental forcing\textemdash are common in drylands, particularly across ecotones between (semi)-arid vegetated areas and deserts, where self-organized patterns are more common \cite{barbier2006self,Deblauwe2011}. Incorporating both of these features into dryland models challenges the prevailing interpretation that self-organized vegetation patterns indicate enhanced ecosystem resilience \cite{Siteur2014,Rietkerk2021,Tarnita2024,Moreno2025}. 

\begin{figure*}[t!]
	\centering
	\includegraphics[width=0.85\textwidth]{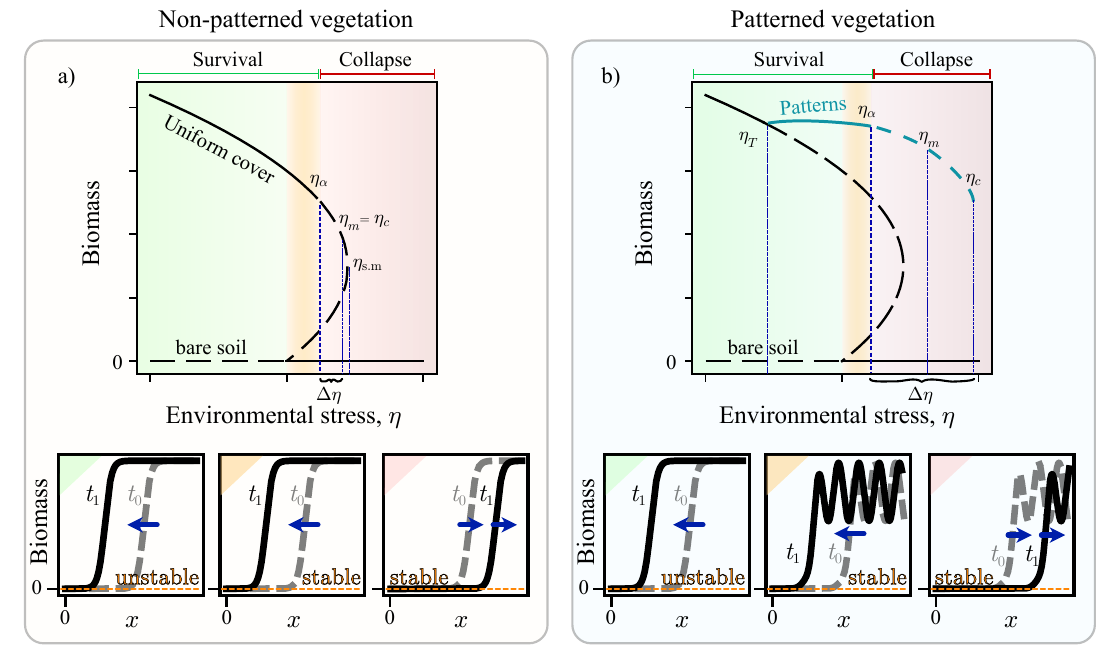}
	\caption{Schematic summary of the effect of non-reciprocal plant interactions in homogeneous (a) and patterned (b) ecosystems. In both scenarios, \ed{nonlinear} convective instabilities anticipate ecosystem collapse relative to the corresponding isotropic tipping point $\eta_{\alpha}<\eta_{c}$. This shift is stronger in patterned vegetation, which can make self-organized ecosystems collapse at stress levels that homogeneous vegetation can withstand.In terms of the vegetation-desert \ed{front} velocity, we distinguish three regimes. (i) At low environmental stress (green region in a, b; $\eta<0$), the bare-soil state $b=0$ is unstable and vegetation fronts can invade desert regions (leftmost panel in the bottom row of a and b). (ii) At intermediate stress levels (orange region in a, b), the desert state becomes stable but vegetation fronts can still invade deserts. (iii) At high environmental stress  (red region in a, b), the front velocity reverts and triggers a desertification front propagation into the vegetated area, exacerbated by nonreciprocal interactions.}
	\label{F4}
\end{figure*}

Under the more realistic conditions considered in our model, non-reciprocal plant interactions can trigger the propagation of desertification fronts, both in patterned and non-patterned ecosystems, \ed{via a new mechanism that we identified mathematically as a nonlinear convective instability \cite{chomaz1992absolute}. These instabilities require lower levels of environmental stress than isotropic desertification fronts or abrupt tipping. Consequently, ecosystems under increasing environmental stress are likely to become sensitive to nonlinear convective instabilities first, making non-reciprocal interactions a potential key driver of desertification dynamics and reinforcing the ecological relevance of these instabilities. Importantly, spatial patterning amplifies their effects, causing self-organized ecosystems to collapse under stress levels that uniform vegetation can withstand (Fig.\,\ref{F4}).}

Desertification by front propagation is also possible in isotropic ecosystems where plants interact reciprocally, and both in patterned and homogeneous vegetation \cite{bel2012gradual,meron2015ModelNew,zelnik2018regime}. In these conditions, however, spatial patterning tends to enhance resilience by slowing down\textemdash and in some cases even stopping \cite{clerc2012origin}\textemdash front propagation. \ed{We recover this behavior when considering reciprocal interactions, as non-patterned ecosystems can develop desertification fronts when environmental stress crosses the Maxwell point, and front velocities connecting the vegetated region with the surrounding desert reverse. In contrast, patterned ecosystems remain stable in those conditions because they consist of a sequence of localized structures that survive beyond the boundary of the pinning region.

The degree of reciprocity in plant interactions is thus key to determining whether spatial self-organization enhances dryland resilience or, conversely, makes ecosystems more vulnerable to desertification. Consequently, even if two ecosystems display similar spatial patterns, they may respond differently to increasing environmental stress depending on whether their underlying individual-level interactions are reciprocal or not. This difference highlights the difficulties in extracting system-level properties from observed patterns \cite{Tarnita2024} and the importance of incorporating fine-scale environmental anisotropies—along with their effects on plant-plant interactions—into dryland models.}

\ed{Quantifying the sensitivity of real ecosystems to convective instabilities is, however, a challenging task. Single-time snapshots can reveal environmental anisotropy \cite{Deblauwe2011,pinto2023topological}, but might be insufficient to evaluate ecosystem susceptibility to these instabilities. In our reduced model, the Turing instability leading to pattern formation is independent of anisotropy, which mainly affects front velocities. Consequently, accurately determining anisotropy levels\textemdash and the associated desertification risk from fronts driven by non-reciprocal interactions\textemdash requires spatiotemporal datasets or direct field observations \cite{Trautz2017}. In more complex setups, environmental anisotropies can be inferred by jointly analyzing remotely sensed vegetation data and other ecosystem features, such as topography \cite{gandhi2018topographic,mcgrath2012microtopography,Florinsky1996}, wind, and directional fog \cite{Stanton2014,hidalgo2024nonreciprocal}. Models accounting for such environmental complexity at these different scales will provide more reliable insights into how vegetation spatial dynamics influence ecosystem resilience \cite{bonachela2015termite,pintoramos2024arxiv}.}

Another feature that makes desertification via \ed{nonlinear} convective instabilities particularly relevant is that they arise in various modeling frameworks and within broad parameter regimes, including the weak-anisotropy regime discussed earlier. Additionally, patterned ecosystems only require weak non-reciprocity to become more sensitive to convective instabilities than homogeneous vegetation covers, which suggests that patterns could indicate ecosystem weakness under a broad range of environmental conditions. The model independence of our results, provided that they can be reduced to the simplified equation, is also particularly relevant given the diversity of available models and their lack of empirical validation \cite{MartinezGarcia2023}. Our results thus question the prevailing view that self-organized patterns enhance ecosystem resilience \cite{Siteur2014,Rietkerk2021,Tarnita2024,Moreno2025}. Recent results have also highlighted this possibility, showing that within relatively narrow parameter ranges, self-organized patterns can be transient and unstable, leading to ecosystem collapse rather than preventing it \cite{van2025vegetation}.

In summary, our work highlights the importance of isotropy-breaking heterogeneities and desert-dryland interfaces in determining ecosystem resilience and, more generally, in spatial ecological dynamics. Smaller-scale spatial heterogeneities are thought to lead to more complex spatial patterns \cite{echeverria2023effect,bonachela2015termite} and increase ecosystem resilience \cite{pintoramos2024arxiv,rietkerk2021evasion}, while we have shown that constant environmental gradients reduce ecosystem resilience, particularly in the presence of vegetation patterns. Future research should combine mathematical modeling with data-driven parameterizations of spatial heterogeneities, accounting for landscape features, environmental covariates, and anthropogenic pressures \ed{in two-dimensional frameworks} \cite{Florinsky1996, barbier2006self}. This approach would fully characterize how spatial dynamics driven by the intrinsic scales of the water-vegetation feedback interact with endogenous spatial features and determine ecosystem resilience.

\ed{ 
\section*{Materials and Methods}\label{sec:MatMet}

\textbf{Derivation of the reduced equation.} We applied the method of scale separation, reducing the dynamics to the center manifold for the bifurcation of the bare-soil state \cite{elphick1987simple}. Near the onset of this bifurcation, spatio-temporal changes in the variables occur over slow and large temporal and spatial scales, allowing for a systematic reduction of the original models into a reduced equation that only retains leading-order terms. A detailed derivation is given in Supporting Information App. A.
\newline

\noindent \textbf{Numerical simulations.} Spatial terms were treated with centered finite differences; simulations were performed with $dx=0.5$ and $N=682$ points unless otherwise stated. For temporal integration, we employed the fifth-order Dormand-Prince method with adaptive time-step (DOPRI5 in the Scipy Python library), see \cite{pinto2025code:how} for our code. The parameters $\kappa=0.3$ and $d=0.3$ remained fixed throughout all the analyses, but varying them does not qualitatively affect our results.
\newline

\noindent \textbf{Bifurcation diagrams.} We obtained the branches of stable equilibria using a continuation-like method with the following steps: (1) For a given set of parameters $(\Gamma, \alpha)$, a simulation is initialized at $\eta=-0.05$ with a random initial condition. (2) The simulation is performed until the system reaches equilibrium, which we defined using a tolerance criterion, $|b(x, t+dt)-b(x,t)|< dt \cdot \text{Th}$ for all $x$ with tolerance threshold $\text{Th}=10^{-5}$ and time discretization $dt=0.01$. (3) Once equilibrium is reached and the simulation is paused, we increase $\eta$ by an amount $d\eta\approx 0.004$, perturb the biomass with a uniform random variable of amplitude $0.01$, and go back to step (2). This protocol is stopped once $\eta=0.2$ is reached. The solutions for each $\eta$ value are used to obtain the maximum in the bulk vegetation biomass, $b_\mathrm{\tiny{max}}$, where the bulk is defined as the simulation domain excluding the first and last 100 points. This procedure captures the equilibrium value of $b_\mathrm{\tiny{max}}$ under a quasi-static change of $\eta$, giving an approximation to the branch of stable equilibria of the bifurcation diagram. 
\newline

\noindent \textbf{Front velocities from numerical simulations.} We obtained front velocities for each parameter combination independently. For each simulation, we considered an initial condition mimicking a vegetation patch centered in the simulation domain and surrounded by bare soil
\begin{eqnarray*}
     b(x, t=0) &= &\tanh \left({x-L/2+\Delta/2}\right) - \\
     &&\tanh\left({x-L/2-\Delta/2}\right),
\end{eqnarray*}
where $\Delta$ is the width of the vegetation patch. We used $L=682$, $dx=0.5$, and $\Delta=100$. This initial condition favors the creation of two fronts
connecting the bare soil with the vegetation cover equilibrium (patterned or non-patterned). Upon entering the regime of stable localized structures, this initial condition promotes the formation of several such structures in the center of the simulation domain. These structures typically interact via repulsive forces between their centers \cite{berrios2020repulsive,clerc2010colonies}. 

To compute the velocity of each front, we tracked its position $i_f(t)$ over the interval $t_i = 1000$ to $t_f = 4500$, and calculated the mean velocity as
\begin{equation}
    v = \frac{i_f(t_f) - i_f(t_i)}{t_f - t_i}.
\end{equation}
We chose $t_i \neq 0$ to exclude the initial transient phase before the system converges to the front solution, and selected $t_f$ sufficiently short to avoid boundary effects. 
The front position $i_f(t)$ was determined as the spatial node where the vegetation biomass $b(i,t)$ crossed the threshold value $b_{\mathrm{th}} = 0.01$. 
Specifically, we defined $i_f$ as the first grid point satisfying
\begin{equation}
    [b(i_f, t) - b_{\mathrm{th}}][b(i_f + 1, t) - b_{\mathrm{th}}] \le 0,
\end{equation}
where \textit{first} refers to scanning from the left (increasing $i$) when identifying the left front and from the right (decreasing $i$) when identifying the right front. By convention, we defined positive velocities correspond to fronts propagating to the right and negative velocities to the left.
\newline

\noindent \textbf{Effect of non-reciprocities on the front velocity.} To develop an analytical argument that explains the qualitative effect of $\alpha$ on the front velocities, we rewrite Eq. \eqref{eq1} as
\begin{eqnarray}
    \frac{\partial b}{\partial t}= f_\text{reciprocal}(b)- \alpha b\frac{\partial b}{\partial x}.\label{eq:rewritten}
\end{eqnarray}
In the case $\alpha=0$, two symmetric, counterpropagating front solutions exist: $b_l=b_f(x-v_lt)$, and $b_r=b_f(-x-v_rt)$. $(l,r)$ denote leftward and rightward propagation, with $v_l=-v_r$. We further assume $\alpha \ll 1$, and that $\alpha$ only modifies front velocity while leaving its shape unchanged. Under these assumptions, $\partial_t b_{(l,r)} = b_{(l,r)}'(-v_{(l,r)})$ and $b_l'=-b_r'$, where $(\cdot)'$ denotes the derivative with respect to the argument. Following standard procedures for calculating a front velocity as a function of its shape \cite{pismen2006patterns}, we substitute the front ansatz into Eq.\,\eqref{eq:rewritten}, multiply both sides by $b_{(l,r)}'$, and integrate over $x$ to obtain
\begin{eqnarray}
    -m_0v_{(l,r)}= \int f_\text{reciprocal}(b_{(l,r)})b_{(l,r)}'dx-  \nonumber \\
    \alpha \int b_{(l,r)}\left(b_{(l,r)}'\right)^2 dx,
\end{eqnarray}
where $m_0 = \int (b_{l}')^2 dx>0$. Next, we can determine the sign of each contribution to the velocity to obtain the qualitative effect of the non-reciprocal term. For $\alpha=0$, $v_l=-v_r$ because $f_\text{reciprocal}(b_r)=f_\text{reciprocal}(b_l)$, but $b_l'=-b_r'$. Defining $v_l=-v_0$, for $\alpha\neq0$ the velocities satisfy
\begin{eqnarray}
    v_l&\approx& -v_0+ \alpha v_1, \nonumber \\
    v_r&\approx&v_0+ \alpha v_1,
\end{eqnarray}
with $v_1= \int b_{l}\left(b_{l}'\right)^2 dx/m_0=\int b_{r}\left(b_{r}'\right)^2 dx/m_0>0$ (as the biomass is strictly positive, and the other term is something squared). The non-reciprocity parameter $\alpha$ thus shifts both velocities in the same direction, and the nonlinear convective instability occurs once $\alpha=v_0/v_1$ is satisfied. Note, however, that closed expressions for the velocities as a function of the system parameters can only be obtained if an analytical expression for the fronts, $b_{(l,r)}$ is known. This is, in general, not possible unless further approximations are made.

\section*{Data Availability}

The code used in this work, including the numerical simulation schemes and the data analysis scripts, is publicly available in the Rossendorf Data Repository (RODARE), under the unique digital identifier http://doi.org/10.14278/rodare.3744 \cite{pinto2025code:how}

}
\section*{Acknowledgments}
This work was partially funded by the Center of Advanced Systems Understanding (CASUS), which is financed by Germany’s Federal Ministry of Education and Research (BMBF) and by the Saxon Ministry for Science, Culture and Tourism (SMWK) with tax funds on the basis of the budget approved by the Saxon State Parliament. RMG was also supported by FAPESP through grant ICTP-SAIFR 2021/14335-0.


\newpage

\onecolumn

\appendix

\makeatletter
\renewcommand{\part}[1]{%
  \refstepcounter{part}
  \addcontentsline{toc}{part}{#1}
  \etoc@startlocaltoc
}
\makeatother
\part{Supplementary Material}

\begin{center}\huge Supporting Information for: 'How spatial patterns can lead to less resilient ecosystems'\end{center}
\localtableofcontents
\locallistoffigures

\linespread{1.3}

\setcounter{figure}{0}
\renewcommand{\thefigure}{S\arabic{figure}}
\newpage

\section{Derivation of the reduced model, Eq.\,(1)}\label{app:redeq}
\subsection{Interaction-redistribution, kernel-based models.}
Within this family of models, we consider an extension of the logistic equation that includes the effect of long-range plant competition and facilitation via nonlocal terms \cite{lefever1997origin,Tlidi2008},
\begin{equation}
	\frac{\partial B}{\partial t}=r m_f(\tilde{B}) \left(1-\frac{B}{K}\right) B-  m\, m_c(\tilde{B}) B+D \frac{\partial^2 B}{\partial x^2},
	\label{Eq-IntegrodifferentailModel_0}
\end{equation}
where $B$ is the biomass density field, and $x, t$ are the spatial coordinate and time, respectively. $r$ is the intrinsic growth rate in the absence of feedback and stress; $m$ measures the additional mortality due to environmental stress; and $D_b$ quantifies vegetation spreading by seed dispersal and/or clonal reproduction. Long-range interactions are introduced by the coefficients $m_f(\tilde{B})$ and $m_c(\tilde{B})$, which quantify the intensity of facilitation and competition, respectively. Following \cite{Tlidi2008}, these coefficients depend both on the weighted average biomass density within a range of a focal location, $\tilde{B}$, and an interaction strength factor $\xi_{f,c}$,
\begin{equation}\label{eq:nonloc-intred}
 m_{f,c}(\tilde{B})=\exp\left(\xi_{f,c}\int \phi_{f,c}({x'}) B(x+x',t) dx'\right)   
\end{equation}
where $\phi_{f,c}({x'})$ are normalized kernels defining the strength of the interaction between vegetation biomass at locations $x$ and $x'$. We assume these kernels are normalized Gaussians
\begin{equation}\label{eq:kernel-intred}
    \phi_{f,c}(x)\propto\exp{\left[-\frac{(x-x_{0f,0c})^2}{2l_{f,c}^2}\right]}
\end{equation}
with standard deviations $l_{f}$ and $l_{c}$ defining the characteristic scale of facilitation and competition, respectively. To account for non-reciprocity in plant interactions, we included a shift $x_{0f,0c}$ so the maximum interaction strength is not achieved at the focal location $x$ and right and left neighbors interact differently with vegetation biomass at $x$. This shift thus breaks the reflection symmetry $x \leftrightarrow -x$ in Eq.\,\eqref{Eq-IntegrodifferentailModel_0}. Note that for vanishing stress and feedback strengths, $m=\xi_{f,c}=0$, Eq.\,\eqref{Eq-IntegrodifferentailModel_0} reduces to the Fisher-Kolmogorov-Petrovsky-Piskunov (FKPP) equation.

Because Eq. \eqref{Eq-IntegrodifferentailModel_0} is rather complex due to its nonlinearities and nonlocalities, it is hard to work with it analytically and even numerically. We present a method to derive a simpler equation that retains the behavior of the original one near a critical point in parameter space \cite{elphick1987simple,kozyreff2007nonvariational}. This method relies on performing a nonlinear change of coordinates to the \textit{center manifold} of a bifurcation present in the original equation through the following steps. First, we identify a bifurcation in the original equation (our critical point). At that bifurcation, the system has a vanishing eigenvalue, meaning that the dynamics is arbitrarily slow in the direction of the phase space defined by the corresponding eigenvector.  This separation of time scales will allow us to reduce the dynamical system to this slow direction because all the other directions in the phase space will rapidly relax and become dynamically irrelevant. Second, we propose a polynomial change of variables using a scaled slow time, assuming that we are close to the critical point. Third, we impose a series of additional restrictions to encapsulate a series of transitions and spatiotemporal scales present in the original models, in particular, a nascent bistability, a Turing instability, and a vanishingly small characteristic wavenumber and group velocity. This set of constraints, together with the original restriction of working close to a bifurcation point, allows us to derive a series of self-consistent reduced equations (independent of the distance to the critical point) that capture increasingly complex phenomena--- the more equations we derive, the more complex the phenomena. Lastly, we impose a \textit{solvability condition} on the hierarchy of relationships obtained in the previous step, which allows us to iteratively obtain the nonlinear change of variables up to a desired order. This iterative procedure progressively refines the accuracy of the reduced equation, although only close to the critical point.

We apply this procedure to the nonlocal model in Eq.\,\eqref{Eq-IntegrodifferentailModel_0} to obtain the reduced model we used in the main text. First, we write Eq.\,\eqref{Eq-IntegrodifferentailModel_0} in non-dimensional form by scaling time and the biomass density field
\begin{equation}
   t= \frac{\tau}{r}, \ \ \ \ \ \ \ B= K b.
\end{equation}
and defining scaled parameters
\begin{equation}
    \mu= \frac{m}{r} \ \ \ \ \xi_{f,c}= \frac{\chi_{f,c}}{K}, \ \ \ \ D= \frac{D_B}{r},
\end{equation}
obtaining
\begin{equation}
	\frac{\partial b}{\partial \tau}=m_f(1-b)b-\mu m_c b+D \frac{\partial^2 b}{\partial x^2}.
	\label{Eq-IntegrodifferentailModel}
\end{equation}
Next, we find a critical point of Eq. \eqref{Eq-IntegrodifferentailModel}. To this end, we consider the homogeneous solutions $b_h$ satisfying
\begin{eqnarray}
    0= b_h(1-b_h)e^{\chi_f b_h} - \mu b_h e^{\chi_c b_h}.
\end{eqnarray}
$b_h=0$ is always a solution, and the remaining possible solutions satisfy
\begin{eqnarray}
    \mu e^{-(\chi_f-\chi_c)b_h} = (1-b_h). \label{bistability}
\end{eqnarray} 
Eq.\,\eqref{bistability} defines two curves in the $(b, \mu)$ plane, and the intersections between them correspond to possible equilibria. A bifurcation occurs when the parameters are such that two or more solutions $b_h$ collapse to a single point, for example, to $b_h=0$. The homogeneous solutions of the nonlocal Eq. \,\eqref{Eq-IntegrodifferentailModel} present a bifurcation at 
$$\mu=1$$. Analyzing the linear dynamics, 
\begin{eqnarray}
    b = 0 + \delta e^{ikx + \lambda t}, \nonumber \\
    \lambda = 1-\mu - Dk^2, \label{linearized}
\end{eqnarray}
we can identify this bifurcation with a transcritical bifurcation at which the equilibrium $b_h=0$ changes its stability from being unstable for $\mu<1$ to stable for $\mu>1$. Having identified a bifurcation, we will state the additional conditions that we will impose:
\begin{enumerate}
    \item Eq.\,\eqref{Eq-IntegrodifferentailModel} can show bistability or monostability. We will perform our analysis near the transition from one case to the other to cover both. 
    \item We are also interested in patterns, so we will impose proximity to a Turing instability.
    \item Because we are interested in macroscopic patterns, we focus on the case where patterns exhibit big wavelengths, or equivalently, wavenumbers $k\rightarrow 0$.
    \item Lastly, the nonreciprocal interaction induces a velocity. We require that the timescale of this movement be the same order of magnitude as that of the evolution of the homogeneous or patterned solution.
\end{enumerate}

These four conditions, together with the proximity to the bifurcation point, reduce our analysis to the neighborhood of a single point in a five-dimensional parameter space (four constraints plus the bifurcation condition). This number of conditions defines the codimension of the reduced equation. 

The transition from monostability to bistability can be readily determined by looking for the condition for which, at $b_h=0$, the curves defining the remaining equilibria, Eq. \eqref{bistability}, are tangent to one another. By differentiating \eqref{bistability} with respect to $b_h$ and evaluating at $b_h=0$ for $\mu=1$ one gets
$$\chi_f - \chi_c = 1,$$
which allows us to obtain the order of magnitude of the biomass as a function of the distance to the bifurcation point. Let 
$$ \mu =1 + \epsilon \eta, $$
$$ \chi_f - \chi_c = 1 + \chi,$$
where $\epsilon\ll1$ quantifies the distance to the bifurcation, $\eta \sim O(1)$ is a constant, and $\chi\ll1$ quantifies how far we are from the nascent bistability. Solving for the biomass to the lowest order in Eq. \eqref{bistability} leads to
$$b_h \approx \chi \pm \sqrt{\chi^2 - 2 \epsilon \eta}.$$
The solution $b_h$ is determined from both parameters simultaneously only if $\chi\sim O(\epsilon^{1/2})$. Otherwise, just one of the parameters determines the solution and bistability is lost. This condition leads us to write
$$\chi = \epsilon^{1/2} \kappa,$$
from which it follows that $b\sim O(\epsilon^{1/2})$.

Next, we proceed to formulate the nonlinear change of variables. We expand the parameters $\mu$, $\chi_f$, $\chi_c$ and the field $b$ as
$$ \mu = 1+ \epsilon \eta + ...$$
$$ \chi_f- \chi_c = 1+\epsilon^{1/2}\kappa$$
$$ b= 0 + B+ b(B)^{[2]}+...$$
$$ \dot{B} = a(B)^{[1]} + a(B)^{[2]}+... $$
where $(\cdot)^{[n]}$ means terms of polinomial order $n$ in the change of variables coordinate $B$. The linear dynamics \eqref{linearized} provides the relevant timescale, $t_0=1/\lambda \sim O(\epsilon^{-1})$, and the relevant spatial scale, $l_0=1/k\sim O( [D/\epsilon]^{1/2})$. Let us for the moment define $(D/\epsilon)^{1/2}=\nu^{-1}$, with $\nu\ll1$. Then, it follows that the center manifold direction evolves in the slow variables $T=\epsilon \tau$ and $X= \nu x$, and remembering that $b\sim O(\epsilon^{1/2})$, we write $B=\epsilon^{1/2}A(T, X)$. Note that $(db/dt)^{[m]}= a(B)^{[m]}$ by construction. Additionally, $\dot{B}= \epsilon^{3/2}(dA/dT)$, so, to obtain an equation independent of $\epsilon$, upon replacing our change of variables, we consider terms with a prefactor of $\epsilon^{3/2}$ only. Let us compute the first term by replacing our change of variables in Eq. \eqref{Eq-IntegrodifferentailModel} 
$$a(B)^{[1]}= \left(-\epsilon \eta + \nu^2D\frac{\partial^2}{\partial X^2}\right)B.$$
For the following terms, we will need to expand the nonlinearities in polynomials, such that
\begin{eqnarray*}
    \exp \left( \chi_i \int \phi_i (x')b(x+x')dx'\right) &=& 1+ \left( \chi_i \int \phi_i (x')b(x+x')dx'\right) \\
    &&+\frac{1}{2}\left( \chi_i \int \phi_i (x')b(x+x')dx'\right)^2+ ... \\
    &=&1+ \chi_i\left(b + c_{1,i}\frac{\partial b}{\partial x}+c_{2,i}\frac{\partial^2 b}{\partial x^2}+ ... \right) \\
    &&+ \frac{\chi_f^2}{2}\left(b + c_{1,i}\frac{\partial b}{\partial x}+c_{2,i}\frac{\partial^2 b}{\partial x^2}+ ... \right)^2 + ...
\end{eqnarray*}
with
$$ c_{n,i}=\frac{1}{n!}\int \phi_i(x)x^ndx.$$
By collecting the terms quadratic in $B$, we obtain
\begin{eqnarray*}
    a(B)^{[2]} &=& (\chi_f -\chi_c -1-\epsilon)B^2 + \chi_fB \left( c_{1f}\frac{\partial B}{\partial x}+c_{2f}\frac{\partial^2 B}{\partial x^2}+ ... \right) \\
    && -(1+\epsilon)\chi_cB\left( c_{1c}\frac{\partial B}{\partial x}+c_{2c}\frac{\partial^2 B}{\partial x^2}+ ... \right),
\end{eqnarray*}
where not all the derivatives will be relevant because, in the slow variable, they read $(\partial^n B/\partial x^n) = \epsilon ^{1/2}\nu^n (\partial^n A/ \partial X^n)$ with $\nu \sim O(\epsilon^q)$. We have not specified $q$ yet, but following condition 3 it must be $q>0$. Therefore, at some point in the expansion, the derivatives will have a prefactor with a power of $\epsilon$ greater than $3/2$. 

Finally, the cubic term in $B$ reads
$$ a(B)^{[3]} = -\frac{1}{2}B^3 + O(\epsilon^{3/2}\nu, \epsilon^2).$$
\noindent We have computed the terms of the nonlinear change of variables  $\dot{B} = a(B)^{[1]} + a(B)^{[2]}+ a(B)^{[3]}$. Naturally, terms $a(B)^{[4]}$ will be of higher order due to the derived scaling of $b$, $b\sim O(\epsilon^{1/2})$.  We explicitly write these terms, replacing the expansion near the critical point $\mu=1+\epsilon\eta$ and $\chi_f-\chi_c=1+\epsilon^{1/2}\kappa$, obtaining
\begin{eqnarray*}
    \epsilon^{3/2} \frac{\partial A}{\partial T}&=&\epsilon^{3/2}\left(-\eta A +\kappa A^2-\frac{A^3}{2}\right) + \epsilon^{1/2}\nu^2 D\frac{\partial^2A}{\partial X^2} \\
    &&+\epsilon\nu(\chi_f c_{1f}-\chi_c c_{1c})A\frac{\partial A}{\partial X}+\epsilon \nu^2(\chi_fc_{2f}-\chi_c c_{2c})A\frac{\partial^2A}{\partial X^2}\\
    &&+\epsilon\nu^3(\chi_f c_{3f}-\chi_c c_{3c})A\frac{\partial^3 A}{\partial X^3}+\epsilon \nu^4(\chi_fc_{4f}-\chi_c c_{4c})A\frac{\partial^4A}{\partial X^4}\\
    &&+O(\epsilon^{3/2}\nu, \epsilon^2,\epsilon \nu^5).
\end{eqnarray*}
\noindent From here, we will impose conditions (2-4) to obtained a closed equation in the limit $\epsilon\rightarrow 0$. The remaining conditions we wish to impose are related to the pattern-formation instability and the induced group velocity. Then, we analyze the linearized equation around a non-zero homogeneous state. Thus, we let $A=A_h+ \delta e^{ikX+\lambda T}$ and analyze $\lambda(k)$. The imaginary part fulfills
$$\epsilon^{3/2}\text{Im} \left[ \lambda \right]= \epsilon \nu (\chi_f c_{1f}-\chi_c c_{1c})kA_h - \epsilon \nu^3 (\chi_f c_{3f}-\chi_c c_{3c})k^3A_h,$$
and the real part
\begin{eqnarray*}
    \epsilon^{3/2}\text{Re}\left[ \lambda\right] &=& \epsilon^{3/2}\left(-\eta+2\kappa A_h -\frac{3}{2}A_h^2 \right)-\left(\epsilon^{1/2}\nu^2D+\epsilon\nu^2A_h(\chi_fc_{2f}-\chi_c c_{2c})\right)k^2\\
    &&+ \epsilon \nu^4A_h(\chi_fc_{4f}-\chi_c c_{4c})k^4.
\end{eqnarray*}
The Turing instability may occur when the term proportional to $k^2$ vanishes. In this limit, the only term stabilizing the equation is the one proportional to $k^4$ (provided that it is negative). Then, it follows that
\begin{eqnarray}
     \epsilon \nu^4A_h(\chi_fc_{4f}-\chi_c c_{4c})&\sim& O(\epsilon^{3/2}), \nonumber\\
    \left(\epsilon^{1/2}\nu^2D+\epsilon\nu^2A_h(\chi_fc_{2f}-\chi_c c_{2c})\right) &=& 0 + O(\epsilon^{3/2}).
    \label{Turing_inf_wav}
\end{eqnarray}
Equations \eqref{Turing_inf_wav} are simultaneously solved for $\nu=\epsilon^{1/8}$,
$$D =0+ \epsilon^{3/4}d,$$
and
$$c_{2f}+\chi_c(c_{2f}-c_{2c})= 0 + \epsilon^{1/4}\chi_1.$$
\noindent It is easy to verify that the Turing instability occurs for the characteristic wavevector $k_c=0+ O(\epsilon ^{1/8})$, so we are indeed looking at macroscopic patterns of large wavelength consistent with the slow spatial scale $l_0$. The remaining condition concerns the induced velocity, which is encapsulated in the imaginary part of $\lambda$. We require that the oscillation of the pattern is on the same timescale as its growth rate. The lowest order corresponds to
$$ \epsilon^{1+1/8} (\chi_f c_{1f}-\chi_c c_{1c}),$$
Then, it must be fulfilled that
\begin{equation}
    c_{1f} + \chi_c(c_{1f}-c_{1c})=0+s\epsilon^{3/8}.
    \label{Advection}
\end{equation} 
Regarding the group velocity induced by the non-reciprocity term, we asuume that, because nonreciprocity is caused by a single process, the nonreciprocal coupling parameters, $x_{0f}$ and $x_{0c}$ have a single origin and are, thus, proportional to a single parameter. Let $$x_{0f}=v \alpha_f$$ and $$x_{0c}=v \alpha_c.$$ Then, the Eq. \eqref{Advection} is fulfilled for $v= 0 + \epsilon^{3/8}\alpha$, and it follows that $\epsilon \nu^3 (\chi_f c_{3f}-\chi_c c_{3c})\sim \epsilon^{3/2 + 1/4}$. 

Summarizing, considering a region of parameters near the critical point determined by a bifurcation and the four additional conditions we imposed, with the distance to the critical point measured by the bookkeeping parameter $\epsilon\ll1$, we let 
\begin{eqnarray*}
    \mu&=& 1+ \epsilon \eta, \\
    \chi_f&=&1+\chi_c +  \epsilon^{1/2}\kappa, \\
    \chi_c&=& \frac{l_f^2}{l_c^2-l_f^2}+\epsilon^{1/4}\Gamma,\\
    D&=&0 +\epsilon^{3/4}d,\\
    v&=&0 + \epsilon^{3/8}\alpha,\\
    b&=&0+\epsilon^{1/2}A(T=\epsilon \tau, X=\epsilon^{1/8}x),
\end{eqnarray*}
and insert those expressions in Eq.\,\eqref{Eq-IntegrodifferentailModel}. We obtain the equation for $A(X,T)$ reading
\begin{eqnarray}
    \epsilon^{3/2} \frac{\partial A}{\partial T}&=& \epsilon^{3/2}\left(-\eta A+\kappa A^2-\frac{A^3}{2}+ (d- (l_c^2-l_f^2)\Gamma A)\frac{\partial^2A}{\partial X^2}-3l_f^2l_c^2A\frac{\partial^4A}{\partial X^4} \right. \\
    &&\left.+\alpha\frac{\alpha_fl_c^2-\alpha_cl_f^2}{l_c^2-l_f^2}\frac{\partial A}{\partial X} \right) + O(\epsilon^{3/2 +1/4}).
\end{eqnarray}

In the limit $\epsilon \rightarrow 0$, this equation describes exactly the behavior of the original system. However, for small values of $\epsilon$, it presents corrections $~O(\epsilon^{1/4})$ that prevent the application of the reduced equation when we move in the parameter space away from the critical point. Importantly, all the constants in this reduced equation are determined by the ecologically relevant parameters of the starting model. It is easy to verify that one obtains exactly the equation \eqref{eq1} in the main text by performing a non-dimensionalization of space. That is, using the nondimensional spatial variable $Z$, defined by $X=(3l_f^2l_c^2)^{1/4}Z$, and redefining the parameters of spatial interactions accordingly. Explicitly, this non-dimensionalization leads to
\begin{eqnarray*}
    \epsilon^{3/2} \frac{\partial A}{\partial T}&=& \epsilon^{3/2}\left(-\eta A+\kappa A^2-\frac{A^3}{2}+ \frac{(d- (l_c^2-l_f^2)\Gamma A)}{(3l_f^2l_c^2)^{1/2}}\frac{\partial^2A}{\partial Z^2}-A\frac{\partial^4A}{\partial Z^4} \right. \\
    &&\left.+\frac{\alpha}{(3l_f^2l_c^2)^{1/4}}\frac{\alpha_fl_c^2-\alpha_cl_f^2}{l_c^2-l_f^2}\frac{\partial A}{\partial Z} \right) + O(\epsilon^{3/2 +1/4}).
\end{eqnarray*}

\subsection{Turing-like coupled water-biomass models}

A similar procedure can be performed in models that describe water-vegetation dynamics explicitly using two variables \cite{pintoramos2024arxiv,fernandez2019front,ruiz2020general}. We consider a classical example of this family of models, originally proposed by Meron and coauthors \cite{gilad2004ecosystem,gilad2007mathematical,meron2015ModelNew} as a generalization of Klausmeier's model \cite{klausmeier1999regular}
\begin{eqnarray}
\frac{\partial B}{\partial t} &=& R BW\left(1-\frac{B}{K}\right)(1+EB)^2 - MB + D_B(B) \frac{\partial^2 B}{\partial x^2}, \nonumber \\
\frac{\partial W}{\partial t} &=& P - LW - GBW(1+EB)^2 + D_W\frac{\partial^2 W}{\partial x^2} - V\frac{\partial W}{\partial x}.
\label{meron_model}
\end{eqnarray}
where $B$ and $W$ are the biomass and soil water density as a function of space $x$ and time $t$. In the vegetation equation, $R$ is the biomass growth rate per water density, and $M$ and $E$ are the plant mortality rate root-to-shoot ratio, respectively. The diffusion $D_B(B)$ accounts for plant dispersal, which we consider non-linear both to account for density-dependent dispersal effects and to ensure that the reduced equation is independent of the expansion parameter $\epsilon$. In the soil water equation, $P$ is the precipitation parameter, $L$ the evaporation rate, and $G$ the water absorption per biomass density rate. $D_W$ is the water diffusion rate in the soil, and $V$ modulates the intensity of the water runoff. A range of realistic values for these parameters in a specific ecosystem is provided in  \cite{meron2015ModelNew}. 

To perform the model reduction, we first assume that $D_B = D_{0} + D_{1} B + ...$ and define the following dimensionless variables and parameters,
\begin{align*}
t&= \frac{\tau}{L} , & x&= \sqrt{ \frac{D_W}{L}} z, & B&= Kb,  \\
W&= \frac{L}{R} w, & p&= \frac{PR}{L^2}, & m&= \frac{M}{L}, \\
\delta&= EK, & d_0&= \frac{D_0}{D_W}, & d_1&= \frac{D_1 K}{D_W}, \\
\gamma&=\frac{G K}{L}, & s&= \frac{V}{\sqrt{L D_W}}. & &
\end{align*}

\noindent Using these new quantities, we can rewrite Eqs.\,\eqref{meron_model} as 

\begin{eqnarray}
    \frac{\partial}{\partial \tau} \begin{pmatrix} b \\ w \end{pmatrix} = \begin{pmatrix} bw(1-b)(1+\delta b)^2-mb \\ p-w-\gamma bw(1+\delta b)^2 \end{pmatrix} + \begin{pmatrix} (d_0 + d_1 b) \frac{\partial^2 b}{\partial z^2} \\ \frac{\partial^2 u}{\partial z^2} - s\frac{\partial u}{\partial z} \end{pmatrix}. \label{meron_nondim}
\end{eqnarray}
whose homogeneous solutions satisfy 
\begin{eqnarray*}
    w_h{b_h(1-b_h)(1+\delta b_h)^2}&=& {mb_h},\\
    w_h&=&\frac{p}{1+\gamma b(1+\delta b)^2}. 
\end{eqnarray*}
One solution corresponds to the trivial bare-soil solution
\begin{eqnarray*}
    \begin{pmatrix}
        b_0 \\
        w_0
    \end{pmatrix} = \begin{pmatrix}
        0 \\
        p
    \end{pmatrix}.
\end{eqnarray*}
and the nontrivial solutions correspond to the intersection of the curves
\begin{eqnarray*}
    w_b(b)&=& \frac{m}{(1-b_h)(1+\delta b_h)^2},\\
    w_w(b)&=&\frac{p}{1+\gamma b(1+\delta b_h)^2},
\end{eqnarray*}
in the plane $(w, b)$. We find the nascent bistability by imposing that these equilibria collapse to a single point simultaneously. This is achieved by imposing that
\begin{eqnarray*}
   \left. \frac{d w_b}{d b_h} \right|_{b_h=0}&=& \left. \frac{w_w}{d b_h} \right|_{b_h=0},\\
   -m\left( 2\delta- 1 \right) &=& -p\gamma.
\end{eqnarray*}
As we did for the nonlocal model, we impose this condition at a bifurcation of the $b_0=0$ state. We additionally impose the existence of a Turing instability, that spatial structures are macroscopic ($k\rightarrow0$), and that the velocity of these structures is on the same timescale as the growth-rate of homogeneous and patterned solutions (i.e., the same conditions 1-4 mentioned in the previous section). 

To derive the reduced equation, we write the model using the bare soil solution as the reference state, $w= p +u$, and expand the nonlinearities in polynomials. After these manipulations, the system reads
\begin{eqnarray}
\frac{\partial}{\partial \tau} \begin{pmatrix} b \\ u \end{pmatrix} = \begin{pmatrix} p-m & 0 \\ -\gamma p & -1 \end{pmatrix}\begin{pmatrix} b \\ u \end{pmatrix} + \begin{pmatrix} f(b,u) + (d_0 + d_1 b) \frac{\partial^2 b}{\partial z^2} \\ -\gamma g(b,u) + \frac{\partial^2 u}{\partial z^2} - s\frac{\partial u}{\partial z} \end{pmatrix}, \label{WBmodel}
\end{eqnarray}
where
\begin{eqnarray}
f(b,u) &=& ub + b^2 (2\delta p -p) + b^3(p\delta^2-2p\delta) + ub^2(2\delta-1) + O(b^4, ub^3), \\
g(b,u) &=& ub + b^2 2\delta p + b^3 p\delta^2 +ub^2 2\delta + O(ub^3).
\end{eqnarray}
The linear part of Eq.\,\eqref{WBmodel}
shows a change in the stability of the bare soil $(b,\,u)=(0,0)$ when $p=m$, and the form of the linear part (the Jacobian), one deduces it is a transcritical bifurcation. Letting $$p=m-\epsilon \eta,$$ the slow eigenvalue is $$\lambda=-\epsilon \eta.$$ At the bifurcation point $\epsilon=0$, the tangent to the center manifold is the eigenvector of the Jacobian
\begin{eqnarray*}
    \mathbf{J}= \begin{pmatrix}
        0 & 0 \\
        -\gamma p & -1
    \end{pmatrix},
\end{eqnarray*}
which is 
\begin{eqnarray*}
    \mathbf{v}= C\begin{pmatrix}
        1 \\
        -\gamma p
    \end{pmatrix}.
\end{eqnarray*}
This is our starting point to obtain the change of variables to the center manifold in the neighborhood of the bifurcation and our further conditions. The amplitude along this direction in phase space is the variable we will describe; let us call it $C$. This amplitude will evolve spatio-temporally on a slow timescale---because we are close to a bifurcation---and on a slow spatial scale---because we will impose that the wavevector $k\rightarrow0$. Let these scales correspond to $t_0= 1/\lambda \sim O(\epsilon^{-1})$ and $l_0= 1/k \sim O(\nu^{-1})$, with $\nu$ related to $\epsilon$ but unknown for the moment. We define the spatio-temporal slow variables $T=\epsilon \tau$ and $Z=\nu z$. We note that the nascent bistability condition evaluated at the bifurcation point reads 
$$\gamma=2\delta -1.$$
Again, solving for the homogeneous solutions at lowest order reveals that for bistability to exist, it must be satisfied that
$$\gamma=2\delta -1-\epsilon^{1/2}\kappa,$$
from which it follows that $b_h\sim O(\epsilon^{1/2})$.
With those ingredients, we can write for Eq. \eqref{WBmodel} the following
\begin{eqnarray}
\epsilon \frac{\partial}{\partial T} \begin{pmatrix} b \\ u \end{pmatrix} = \mathbf{J}\begin{pmatrix} b \\ u \end{pmatrix} + \begin{pmatrix} f(b,u) \\ -\gamma g(b,u)  \end{pmatrix}+ \begin{pmatrix}
    -\epsilon \eta b \\
    0
\end{pmatrix} +\begin{pmatrix}
    \nu^2 (d_0 + d_1 b)  \frac{\partial^2 b}{\partial Z^2} \\
     \nu^2\frac{\partial^2 u}{\partial Z^2} - \nu s\frac{\partial u}{\partial Z}
\end{pmatrix}, \label{WB_normal_form}
\end{eqnarray}
and perform the change of variables such that
\begin{eqnarray}
    \begin{pmatrix}
        b \\
        u
    \end{pmatrix}&=& \epsilon^{1/2}A \mathbf{v} + \epsilon^{1/2}\nu\begin{pmatrix}
        b(A) \\
        u(A)
    \end{pmatrix}^{[1/2,1]}+ \epsilon \nu\begin{pmatrix}
        b(C) \\
        u(C)
    \end{pmatrix}^{[1,1]}+...,\nonumber \\
    \epsilon^{3/2}\frac{\partial A}{\partial T}&=& \epsilon^{1/2}a(A)^{[1/2,0]}+\epsilon^{1/2}\nu a(A)^{[1/2,1]}+\epsilon \nu a(A)^{[1,1]}+....
\label{change_of_variables}
\end{eqnarray}
$(\cdot)^{[n, m]}$ are just bookkeeping superscripts to label the functions proportional to $\epsilon^{n}$ and $\nu^{m}$. We insert those expressions into Eq.\,\eqref{WB_normal_form} and solve iteratively the resulting hierarchy, noting that
\begin{eqnarray*}
    \epsilon\frac{\partial}{\partial \tau} \begin{pmatrix} b \\ u \end{pmatrix}^{[n,m]}= \epsilon^{n}\nu^m a(A)^{[n,m]} \mathbf{v}.
\end{eqnarray*}
\noindent The first-order equation in this hierarchy corresponds to the linearized dynamics, reading

{\large \underline{ $O(\epsilon^{1/2}\nu^{0}):$}}
\begin{eqnarray*}
   \epsilon^{1/2} a(A)^{[1/2,0]} \mathbf{v} = \epsilon^{1/2}A\mathbf{J}\mathbf{v},
\end{eqnarray*}
and considering how $\mathbf{J}$ acts on $\mathbf{v}$, we find that $a(A)^{[1/2, 0]}$ vanishes. We solve iteratively for the next order

{\large \underline{ $O(\epsilon^{1/2}\nu^{1}):$}}
\begin{eqnarray*}
    \epsilon^{1/2}\nu a(A)^{[1/2, 1]} \mathbf{v} = \epsilon^{1/2}\nu \mathbf{J}\begin{pmatrix}
        b(A) \\
        u(A)
    \end{pmatrix}^{[1/2,1]} + \epsilon^{1/2}\nu\begin{pmatrix}
        0 \\
        \gamma p s \frac{\partial A}{\partial Z},
    \end{pmatrix}
\end{eqnarray*}
which has to be solved for two unknowns, $a(A)^{[1/2, 1]}$ and the vector of the change of variables $\left((b(A), u(A))^{[1/2,1]}\right)^T$ (where $T$ means the transpose).

Note that the problem is linear, so solutions must fulfill the \textit{solvability condition}. The solvability condition states that the system $\mathbf{A}\mathbf{x}=\mathbf{b}$ has a solution if and only if $\mathbf{b} \perp \text{Ker} \{ \mathbf{A}^{\dagger}\}$, where $(\cdot)^{\dagger}$ means the adjoint, or conjugated transpose under the usual inner product. Therefore, we need to compute the kernel of the operator 
\begin{eqnarray*}
    \mathbf{J}^{\dagger}= \begin{pmatrix}
        0 & -\gamma p \\
        0 & -1
    \end{pmatrix}.
\end{eqnarray*}
which corresponds to
\begin{eqnarray*}
    \text{Ker}\{ \mathbf{J}^\dagger\} = \left\{ \begin{pmatrix}
        1 \\
        0
    \end{pmatrix}\right\}.
\end{eqnarray*}
Next, we impose that 
\begin{eqnarray*}
   \left[ a(A)^{[1/2,1]}\mathbf{v}-\begin{pmatrix}
        0 \\
        \gamma p s \frac{\partial A}{\partial Z}
    \end{pmatrix} \right] \perp \begin{pmatrix}
        1 \\
        0
    \end{pmatrix},
\end{eqnarray*}
or equivalently
\begin{eqnarray*}
   \left[ a(A)^{[1/2,1]}\mathbf{v}-\begin{pmatrix}
        0 \\
        \gamma p s \frac{\partial A}{\partial Z}
    \end{pmatrix} \right] \cdot\begin{pmatrix}
        1 \\
        0
    \end{pmatrix}=0.
\end{eqnarray*}
This implies that
\begin{eqnarray*}
    a(A)^{[1/2,1]}=0,
\end{eqnarray*}
which leads to the equation
\begin{eqnarray*}
    \mathbf{J}\begin{pmatrix}
        b(A) \\
        u(A)
    \end{pmatrix}^{[1/2,1]} = -\begin{pmatrix}
        0 \\
        \gamma p s \frac{\partial A}{\partial Z}
    \end{pmatrix}.
\end{eqnarray*}
This equation has multiple solutions. Nevertheless, the method warrants that the normal form (or the reduced equation) is unique. Then, one must choose the solution such that the change of variables \eqref{change_of_variables} cannot be simplified further with additional changes of variables. We choose
\begin{eqnarray*}
    \begin{pmatrix}
        b(A) \\
        u(A)
    \end{pmatrix}^{[1/2,1]} = \begin{pmatrix}
        0 \\
        \gamma p s \frac{\partial A}{\partial Z}
    \end{pmatrix}.
\end{eqnarray*}
The next order reads

{\large \underline{ $O(\epsilon^{1}\nu^{0}):$}}
\begin{eqnarray*}
    \epsilon a(A)^{[1, 0]} \mathbf{v} = \epsilon \mathbf{J}\begin{pmatrix}
        b(A) \\
        u(A)
    \end{pmatrix}^{[1,0]} + \epsilon\begin{pmatrix}
        0 \\
        -\gamma p A^2
    \end{pmatrix}.
\end{eqnarray*}
Note that we face the same linear problem as before, this is what makes the method iterable up to the desired order. Then, applying the same arguments as in the previos order, we compute the solutions
\begin{eqnarray*}
    a(A)^{[1,0]}&=&0,\\
    \begin{pmatrix}
        b(A) \\
        u(A)
    \end{pmatrix}^{[1,0]} &=& \begin{pmatrix}
        0 \\
        -\gamma p A^2
    \end{pmatrix}.
\end{eqnarray*}
That leads to the next order

{\large \underline{ $O(\epsilon^{1/2}\nu^{2}):$}}
\begin{eqnarray*}
    \epsilon^{1/2}\nu^2 a(A)^{[1/2, 2]} \mathbf{v} = \epsilon^{1/2}\nu^2 \mathbf{J}\begin{pmatrix}
        b(A) \\
        u(A)
    \end{pmatrix}^{[1/2,2]}+ \epsilon^{1/2}\nu^2\begin{pmatrix}
        d_0  \\
        -\gamma p  - \gamma p s^2
    \end{pmatrix}\frac{\partial^2A}{\partial Z^2},
\end{eqnarray*}
with solutions 
\begin{eqnarray*}
    a(A)^{[1/2, 2]}&=& d_0 \frac{\partial^2 A}{\partial Z^2}, \\
    \begin{pmatrix}
        b(A) \\
        u(A)
    \end{pmatrix}^{[1/2,2]} &=& \begin{pmatrix}
        0 \\
        -(d_0 + \gamma p+\gamma p s^2) \frac{\partial^2 A}{\partial Z^2}
    \end{pmatrix}.
\end{eqnarray*}
At the next order, we find

{\large \underline{ $O(\epsilon^{1/2}\nu^{3}):$}}
\begin{eqnarray*}
    \epsilon^{1/2}\nu^3 a(A)^{[1/2, 3]} \mathbf{v} = \epsilon^{1/2}\nu^3 \mathbf{J}\begin{pmatrix}
        b(A) \\
        u(A)
    \end{pmatrix}^{[1/2,3]}+ \epsilon^{1/2}\nu^3\begin{pmatrix}
        0  \\
        \gamma p s  + s(d_0+ \gamma p + \gamma p s^2)
    \end{pmatrix}\frac{\partial^3A}{\partial Z^3},
\end{eqnarray*}
with solutions
\begin{eqnarray*}
    a(A)^{[1/2, 3]}&=& 0, \\
    \begin{pmatrix}
        b(A) \\
        u(A)
    \end{pmatrix}^{[1/2,3]} &=& \begin{pmatrix}
        0 \\
         [\gamma p s  + s(d_0+ \gamma p + \gamma p s^2)] \frac{\partial^3 A}{\partial Z^3}
    \end{pmatrix}.
\end{eqnarray*}
Next

{\large \underline{ $O(\epsilon^{1/2}\nu^{4}):$}}
\begin{eqnarray*}
    \epsilon^{1/2}\nu^4 a(A)^{[1/2, 4]} \mathbf{v} = \epsilon^{1/2}\nu^4 \mathbf{J}\begin{pmatrix}
        b(A) \\
        u(A)
    \end{pmatrix}^{[1/2,4]}+ \epsilon^{1/2}\nu^4\begin{pmatrix}
        0  \\
        -\gamma p s^2  - s^2(d_0+ \gamma p + \gamma p s^2)- (d_0 + \gamma p+\gamma p s^2)
    \end{pmatrix}\frac{\partial^4A}{\partial Z ^4},
\end{eqnarray*}
with solutions
\begin{eqnarray*}
    a(A)^{[1/2, 4]}&=& 0, \\
    \begin{pmatrix}
        b(A) \\
        u(A)
    \end{pmatrix}^{[1/2,4]} &=& \begin{pmatrix}
        0 \\
          -\gamma p s^2  - s^2(d_0+ \gamma p + \gamma p s^2)- (d_0 + \gamma p+\gamma p s^2) 
    \end{pmatrix}\frac{\partial^4 A}{\partial Z^4}.
\end{eqnarray*}
Next

{\large \underline{ $O(\epsilon^{1}\nu^{1}):$}}
\begin{eqnarray*}
    \epsilon \nu a(A)^{[1, 1]} \mathbf{v} = \epsilon\nu \mathbf{J}\begin{pmatrix}
        b(A) \\
        u(A)
    \end{pmatrix}^{[1,1]}+ \epsilon\nu\begin{pmatrix}
        \gamma p s A\frac{\partial A}{\partial Z}  \\
        -\gamma^2 p s A\frac{\partial A}{\partial Z} + 2\gamma p s A\frac{\partial A}{\partial Z} 
    \end{pmatrix},
\end{eqnarray*}
with solutions
\begin{eqnarray*}
a(A)^{[1, 1]}&=& \gamma p s A \frac{\partial A}{\partial Z}, \\
    \begin{pmatrix}
        b(A) \\
        u(A)
    \end{pmatrix}^{[1,1]} &=& \begin{pmatrix}
        0 \\
        \gamma ^2 p^2s - \gamma^2 p s + 2\gamma p s
    \end{pmatrix}A\frac{\partial A}{\partial Z}.
\end{eqnarray*}
Next

{\large \underline{ $O(\epsilon^{1}\nu^{2}):$}}
\begin{eqnarray*}
    \epsilon \nu^2 a(A)^{[1, 2]} \mathbf{v} = \epsilon\nu^2 \mathbf{J}\begin{pmatrix}
        b(A) \\
        u(A)
    \end{pmatrix}^{[1,2]}+ \epsilon\nu^2\begin{pmatrix}
        [-(d_0+\gamma p + \gamma p s^2) + d_1]A\frac{\partial ^2 A}{\partial Z^2} \\
        \gamma (d_0 +\gamma p + \gamma p s)A\frac{\partial ^2 A}{\partial Z^2} -\gamma p \frac{\partial^2 A^2}{\partial Z^2}-s( \gamma ^2 p^2s - \gamma^2 p s + 2\gamma p s)\frac{\partial (A\frac{\partial A}{\partial Z})}{\partial Z}
    \end{pmatrix},
\end{eqnarray*}
with solutions
\begin{eqnarray*}
a(A)^{[1, 2]}&=& [-(d_0+\gamma p + \gamma p s^2) + d_1]A\frac{\partial ^2 A}{\partial Z^2}, \\
    \begin{pmatrix}
        b(A) \\
        u(A)
    \end{pmatrix}^{[1,2]} &=& \begin{pmatrix}
        0 \\
        \left(-\gamma p[-(d_0+\gamma p + \gamma p s^2) + d_1]A\frac{\partial ^2 A}{\partial Z^2}+\gamma(d_0 +\gamma p + \gamma p s)A\frac{\partial ^2 A}{\partial Z^2}- \right.\\
        \left. \gamma p \frac{\partial^2 A^2}{\partial Z^2}-s( \gamma ^2 p^2s - \gamma^2 p s + 2\gamma p s)\frac{\partial (A\frac{\partial A}{\partial Z})}{\partial Z}\right)
    \end{pmatrix}.
\end{eqnarray*}
Next

{\large \underline{ $O(\epsilon^{1}\nu^{3}):$}}
\begin{eqnarray*}
    \epsilon \nu^3 a(A)^{[1, 3]} \mathbf{v} = \epsilon\nu^3 \mathbf{J}\begin{pmatrix}
        b(A) \\
        u(A)
    \end{pmatrix}^{[1,3]}+ \epsilon\nu^3\begin{pmatrix}
       [\gamma p s  + s(d_0+ \gamma p + \gamma p s^2)]A \frac{\partial^3 A}{\partial Z^3} \\
       -\gamma [\gamma p s  + s(d_0+ \gamma p + \gamma p s^2)]A \frac{\partial^3 A}{\partial Z^3}+ \frac{\partial ^2 u(A)^{[1,1]}}{\partial Z^2}-s\frac{\partial u(A)^{[1,2]}}{\partial Z}
    \end{pmatrix},
\end{eqnarray*}
where the terms $u(A)^{[1,1]}$ and $u(A)^{[1,2]}$ have been previously computed and are cumbersome to write (nevertheless, they will have no impact at the end). The solution is
\begin{eqnarray*}
a(A)^{[1, 3]}&=& [\gamma p s  + s(d_0+ \gamma p + \gamma p s^2)]A \frac{\partial^3 A}{\partial Z^3}, \\
    \begin{pmatrix}
        b(A) \\
        u(A)
    \end{pmatrix}^{[1,3]} &=& \begin{pmatrix}
        0 \\
        \left( -\gamma p [\gamma p s  + s(d_0+ \gamma p + \gamma p s^2)]A \frac{\partial^3 A}{\partial Z^3} - \right. \\
        \left. \gamma [\gamma p s  + s(d_0+ \gamma p + \gamma p s^2)]A \frac{\partial^3 A}{\partial Z^3}+ \frac{\partial ^2 u(A)^{[1,1]}}{\partial Z^2}-s\frac{\partial u(A)^{[1,2]}}{\partial Z}\right)
    \end{pmatrix}.
\end{eqnarray*}
Next

{\large \underline{ $O(\epsilon^{1}\nu^{4}):$}}
\begin{eqnarray*}
    \epsilon \nu^4 a(A)^{[1, 4]} \mathbf{v} = \epsilon\nu^4 \mathbf{J}\begin{pmatrix}
        b(A) \\
        u(A)
    \end{pmatrix}^{[1,4]}+ \epsilon\nu^4\begin{pmatrix}
      [ -\gamma p s^2  - s^2(d_0+ \gamma p + \gamma p s^2)- (d_0 + \gamma p+\gamma p s^2) ] A \frac{\partial^4 A}{\partial Z^4} \\
       ...
    \end{pmatrix},
\end{eqnarray*}
We can get the solution for the dynamics of the center manifold coordinate
\begin{eqnarray*}
    a(A)^{[1,4]}= [ -\gamma p s^2  - s^2(d_0+ \gamma p + \gamma p s^2)- (d_0 + \gamma p+\gamma p s^2) ] A \frac{\partial^4 A}{\partial Z^4},
\end{eqnarray*}
The correction to the change of variables will not be needed, thus we do not write it explicitly.
The last relevant contribution (we will check this a posteriori) will be the next order

{\large \underline{ $O(\epsilon^{3/2}\nu^{0}):$}}
\begin{eqnarray*}
    \epsilon^{3/2} a(A)^{[3/2, 0]} \mathbf{v} = \epsilon^{3/2} \mathbf{J}\begin{pmatrix}
        b(A) \\
        u(A)
    \end{pmatrix}^{[3/2,0]} + \epsilon^{3/2}\begin{pmatrix}
        -\eta A + \kappa p A^2+p(4\delta - 3\delta^2)A^3\\
        ...
    \end{pmatrix},
\end{eqnarray*}
and the correction to the dynamics 
\begin{eqnarray*}
    a(A)^{[3/2,0]}= -\eta A + \kappa p A^2+p(4\delta - 3\delta^2)A^3.
\end{eqnarray*}
So far, we have found that
\begin{eqnarray*}
    \epsilon^{3/2}\frac{\partial A}{\partial T}&=& \epsilon^{3/2}\left( -\eta A + \kappa p A^2+p(4\delta - 3\delta^2)A^3\right)+ \epsilon^{1/2}\nu^2d_0\frac{\partial^2 A}{\partial Z^2}+\epsilon \nu \gamma psA\frac{\partial A}{\partial Z}\\
    &&+\epsilon \nu^2 [-(d_0+\gamma p + \gamma p s^2) + d_1]A\frac{\partial ^2 A}{\partial Z^2} +\epsilon \nu^3 [\gamma p s  + s(d_0+ \gamma p + \gamma p s^2)]A \frac{\partial^3 A}{\partial Z^3} \\
    &&+\epsilon \nu^4 [ -\gamma p s^2  - s^2(d_0+ \gamma p + \gamma p s^2)- (d_0 + \gamma p+\gamma p s^2) ] A \frac{\partial^4 A}{\partial Z^4}+ O(\epsilon^{3/2}\nu, \epsilon \nu^5, \epsilon^2).
\end{eqnarray*}
Then, similarly as the case of the nonlocal model, one imposes the conditions for the Turing instability at vanishing wavenumber and the oscilation frequency to be of the same order as the grow rate. One finds that the conditions are satisfied for
\begin{eqnarray*}
    \nu&=& \epsilon^{1/8},\\
    d_0&=& \epsilon^{3/4}d,\\
    s&=&\epsilon^{3/8}\alpha, \\
    d_1&=&\gamma p -\epsilon^{1/4}\Gamma.
\end{eqnarray*}
Finally, one finds the equation
\begin{eqnarray}
    \epsilon^{3/2}\frac{\partial A}{\partial T}&=&\epsilon^{3/2}\left(-\eta A+p\kappa A^2 +p(4\delta -3\delta^2)A^3 + d\frac{\partial^2A}{\partial Z^2}\right. \nonumber\\
    &&\left.+ \alpha\gamma pA\frac{\partial A}{\partial Z}- \Gamma A\frac{\partial ^2 A}{\partial Z^2}-\gamma p A\frac{\partial^4A}{\partial Z^4} \right) + O(\epsilon^{3/2+1/8}).
    \label{reduced_WB}
\end{eqnarray}
Summarizing, the critical conditions and the change of variables are as follows
\begin{eqnarray*}
p&=&m - \epsilon \eta, \\
\gamma&=& 2\delta -1 - \epsilon^{1/2}{\kappa}, \\
d_1&=& \gamma p - \epsilon^{1/4}\Gamma, \\
s&=&  \epsilon^{3/8}\alpha, \\
d_0&=& \epsilon^{3/4}d,\\
\begin{pmatrix}
    b\\
    u
\end{pmatrix}&=& \epsilon^{1/2} A(T=\epsilon t, Z=\epsilon^{1/8}z)\begin{pmatrix}
    1\\
    -\gamma p
\end{pmatrix}+ \begin{pmatrix}
    0 \\
   -\epsilon^{3/4} \gamma p \frac{\partial ^2 A}{\partial Z^2} -\epsilon \gamma p A^2  -\epsilon \gamma p \frac{\partial^4A}{\partial Z^4}+ \epsilon \gamma p \alpha \frac{\partial A}{\partial Z}
\end{pmatrix},
\end{eqnarray*}
inserting them in Eq. \eqref{WB_normal_form},
one finds Eq. \eqref{reduced_WB}, which correspods to the main text equation up to a reescaling of the spatial coordinate and a redefinition of the corresponding parameters. Explicitly, one lets $A= H/[2p|4\delta - 3\delta^2|]^{1/2}$ and $Z=Y(\gamma p)^{1/4}/[2p|4\delta - 3\delta^2|]^{1/8}$, obtaining
\begin{eqnarray*}
    \epsilon^{3/2}\frac{\partial H}{\partial T}&=&\epsilon^{3/2}\left(-\eta H+\frac{p\kappa}{[2p|4\delta - 3\delta^2|]^{1/2}} H^2 -\frac{1}{2}H^3 + \frac{d[2p|4\delta - 3\delta^2|]^{1/4}}{(\gamma p)^{1/2}}\frac{\partial^2H}{\partial Y^2}\right. \nonumber\\
    &&\left. + \frac{\alpha\gamma p[2p|4\delta - 3\delta^2|]^{1/8}}{(\gamma p)^{1/4}}H\frac{\partial H}{\partial Y}- \frac{\Gamma [2p|4\delta - 3\delta^2|]^{1/4}}{(\gamma p)^{1/2}} H\frac{\partial ^2 H}{\partial H^2}-H\frac{\partial^4H}{\partial Y^4} \right) + O(\epsilon^{3/2+1/8}),
    \label{}
\end{eqnarray*}
which corresponds exactly to the main text equation. The corrections are of the $\epsilon^{1/8}$ order when away from the bifurcation point, setting the region of validity of the reduced equation.

\newpage 

\section{The phenomenon in the original models}\label{app:phenmodels}
\begin{figure}[]
    \centering
    \includegraphics[width=0.8\linewidth]{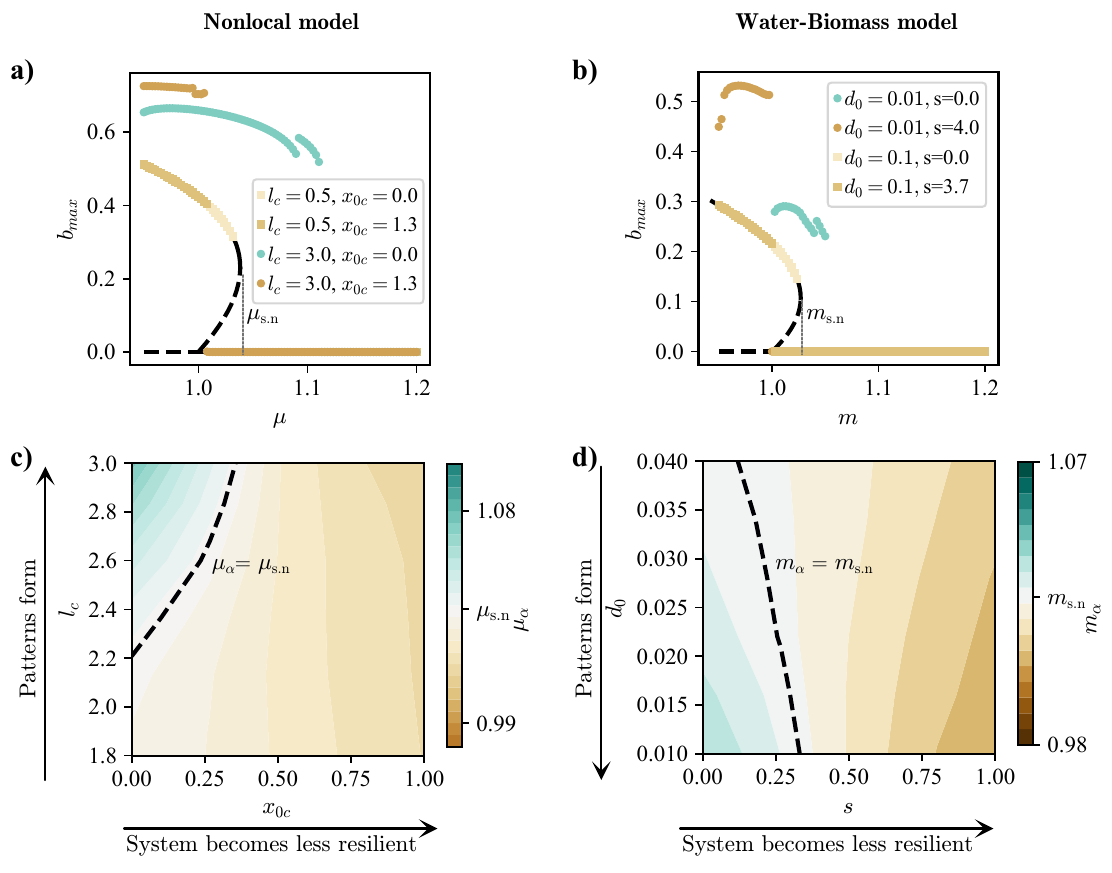}
    \caption[Bifurcation diagram and convective threshold in the original models]{Convective instabilities and ecosystem collapse in the nonlocal (a,\,c) and water–vegetation (b,\,d) models. (a,b) Bifurcation diagrams showing homogeneous (squares) and patterned (circles) vegetation. Colors denote the distance between the non-spatial tipping point $\eta_{s.n}$ and the maximum tolerable stress. Teal (brown) indicates higher (lower) resilience. (c,d) Stress level (relative to $\eta_{s.n}$) at which convective instabilities appear versus pattern-forming parameters ($l_c$, $d_0$) and nonreciprocal strengths ($x_{0c}$, $s$). Parameters. Nonlocal: $\chi_f=3.3$, $\chi_c=2$, $D=0.3$, $l_f=1$, $x_{0f}=0$. Water–biomass: $p=1$, $\delta=1$, $\gamma=0.5$, $d_1=0.001$.}
    \label{FS1}
\end{figure}

For completeness, we performed numerical simulations of the original models we used to derive the reduced equation (Fig.\,\ref{FS1}). Panels a) and b) show typical bifurcation diagrams for the stability of the vegetated solution in the reaction-diffusion water-vegetation, Eqs.\,\eqref{Eq-IntegrodifferentailModel_0}-\eqref{eq:kernel-intred}, and nonlocal interaction-redistribution model, Eq.\,\eqref{meron_model}. In both panels, square markers correspond to non-pattern-forming cases (quasi-homogeneous vegetation) and circles correspond to pattern-forming cases. Increasing the parameter responsible for nonreciprocal interactions\textemdash $x_{0c}$ in the nonlocal model, and $s$ in the water-biomass one\textemdash the vegetation in the bulk dies at levels of environmental stress\textemdash $\mu$ in the nonlocal model and $m$ in the water-biomass\textemdash the ecosystem with reciprocal interactions can withstand. 
Panels c) and d) show the value of the environmental stress at which a nonlinear convective instability arises relative to the tipping point in the non-spatial limit. We compute this environmental threshold for different values of the parameter responsible for the pattern formation instability\textemdash $l_c$ in the nonlocal model and $d_0$ in the water-biomass\textemdash and the non-reciprocity parameter\textemdash $x_{0c}$ in the nonlocal model and $s$ in the water-biomass model. These numerical simulations in the original models confirm that the two key results we obtained with the reduced equation are not an artifact introduced during model reduction. First, nonreciprocal interactions reduce ecosystem resilience for a broad range of parameters. Second, ecosystems exhibiting spatial patterns are more sensitive to convective instabilities than their homogeneous counterparts when non-reciprocity is sufficiently intense. That is, patterned ecosystems can collapse due to nonlinear convective instabilities at levels of environmental stress that homogeneous vegetation withstands. 

\newpage

\section{Convective threshold for smooth mortality gradients}

\begin{figure}[]
    \centering
    \includegraphics[width=0.8\linewidth]{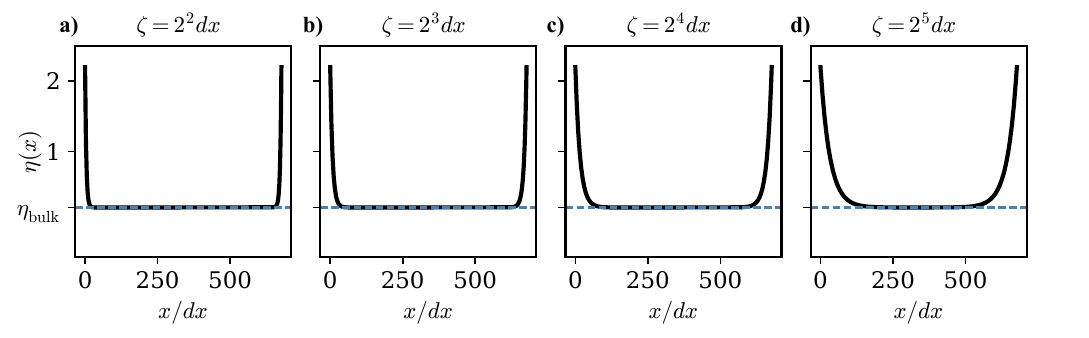}
        \caption[Smooth mortality gradients]{Examples of the smooth mortality gradient toward the boundaries, Eq. \eqref{eq:mortality}. The case $\eta_\text{edge}=2$ is depicted, with different steepness values. (a) $\zeta/dx=2^2$. (b) $\zeta/dx=2^3$. (c) $\zeta/dx=2^4$. (d) $\zeta/dx=2^5$.}
    \label{FS2}
\end{figure}

Our framework relaxes two of the main assumptions that are common in models of vegetation pattern formation in arid environments by accounting for the finite size of vegetated patches and potential anisotropies. In the main text, we account for the finite size of vegetation patches by imposing boundary conditions that set $b=0$ at the patch edges. This modeling choice assumes that the environmental stress (defined in terms of the mortality rate in our model) jumps abruptly from a finite value $\eta$ within the patch to infinity outside it. In this section, we test the sensitivity of our results to the particular choice of the mortality rate, considering a more realistic mortality profile in which $\eta$ increases smoothly and continuously toward the patch edges. Specifically, we used a space-dependent mortality
\begin{eqnarray}
    \eta(x=i \mathrm{d}x) = \eta_{\mathrm{edge}} ( e^{-i/\zeta} + e^{(i-N-1)/\zeta}) + \eta_{\mathrm{bulk}},\label{eq:mortality}
\end{eqnarray}
where $N$ is the total number of nodes used to discretize the space and $\zeta$ sets the steepness of this mortality gradient. $\eta_{\mathrm{edge}}>\eta_{\mathrm{bulk}}$, and hence $\eta(x)$ takes a smaller $\eta_{\mathrm{bulk}}$ within most of the vegetated patch and grows exponentially until it reaches $\eta_{\mathrm{edge}}+\eta_{\mathrm{bulk}}$ at the system boundaries (Fig.\,\ref{FS2}). This mortality profile recovers the Dirichlet boundary conditions used in the main text when $\eta_{\mathrm{edge}}\rightarrow \infty$ and $\zeta\rightarrow 0$. In this scenario, the biomass at the boundary is determined by the dynamics itself, and one can not impose Dirichlet boundary conditions. Therefore, consistent boundary conditions for the numerical integration would be free boundary conditions, which require imposing that all the odd spatial derivatives of $b$ must vanish, or, equivalently, mirroring $b$ at the boundaries.
For this mortality profile and boundary conditions, we computed the bifurcation diagram using the same protocol described in the Methods section of the main text. Provided that $\zeta$ is small enough to make the mortality rate decay sufficiently fast to take the value $\eta_{\mathrm{bulk}}$ within most of the vegegated patch, and $\eta_{\mathrm{edge}}$ is high enough to make the bare-soil state the only stable state in the boundaries, our two main results remain for this smooth mortality profile (Fig.\,\ref{FS3}): (i) non-reciprocal interactions make ecosystems collapse at lower environmental stress; (ii) for sufficiently strong non-reciprocitiy, patterned vegetation can collapse at lower values of $\eta_{\mathrm{bulk}}$ than quasi-homogenous vegetation.

The parameters defining the smooth mortality profile have only a weak influence on the bifurcation diagrams and the locations of the tipping points, which could be within the numerical error of our simulations. For quasi-homogeneous vegetation, increasing $\zeta$ shifts the tipping point by an amount on the order of the discretization used to scan the bulk mortality rate, $\Delta \eta_{\mathrm{bulk}} \approx 0.004$ (left column in Fig.\,\ref{FS3}a, b). For patterned vegetation, changing the values of $\zeta$ changes the shape of the curves of the bifurcation diagram, but without affecting the position of the tipping point (right column in Fig.\,\ref{FS2}a, b). We further confirmed that the location of the tipping points does not change with the decay rate of the mortality rate by computing the tipping points due to convective instabilities $\eta_\alpha$ for patterned and homogeneous vegetation and various values of the edge mortality $\eta_{\mathrm{edge}}$ in the $(\zeta, \alpha)$ parameter space (Fig.\,\ref{FS2}c, d).

\begin{figure}[]
    \centering
    \includegraphics[width=0.7\linewidth]{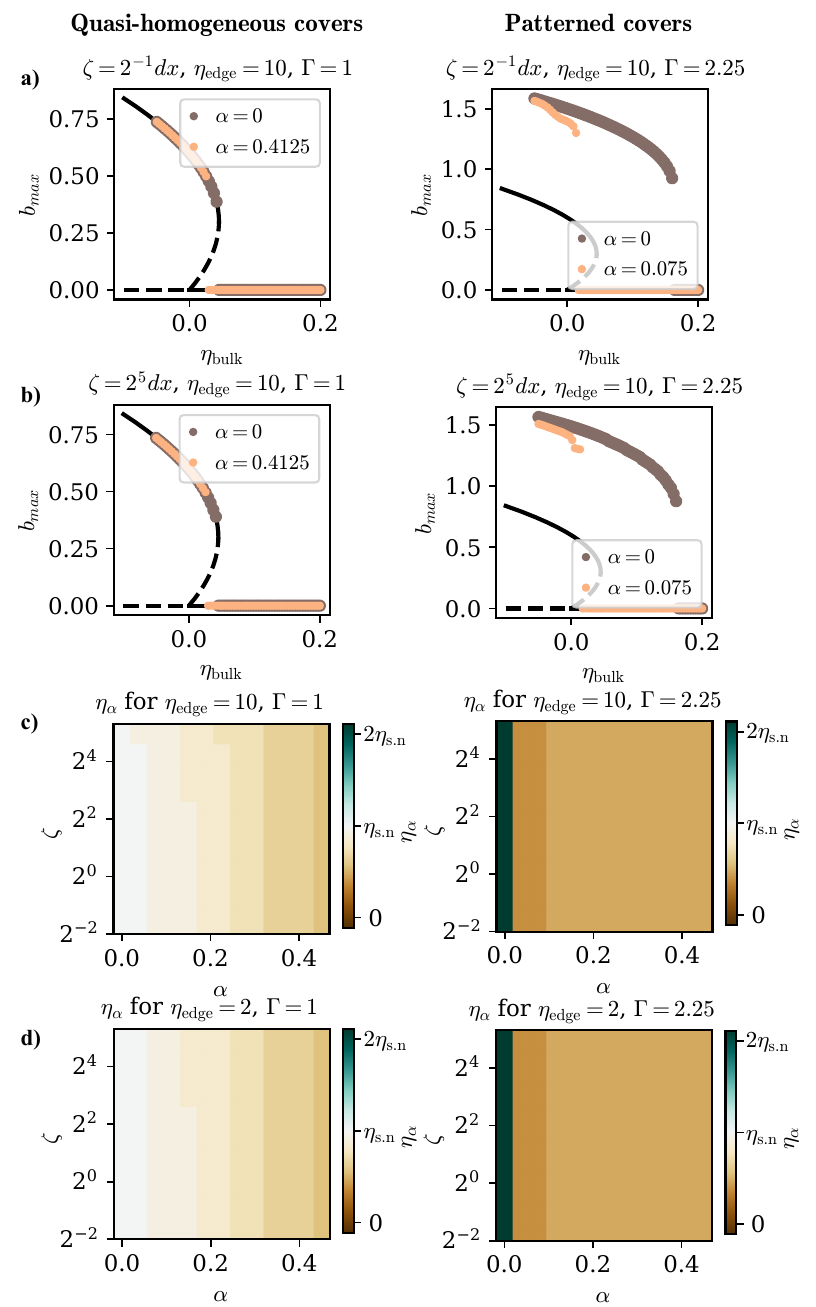}
        \caption[Effect of a smooth mortality gradient]{Effect of a smooth mortality gradient on ecosystem tipping. a,b) Bifurcation diagrams for homogeneous and patterned vegetation with $\eta_{\mathrm{edge}}=10$ and $\zeta/dx=2^{-1}$ (a) or $\zeta/dx=2^{5}$ (b). c,d) Tipping point induced by convective instabilities in the $(\zeta, \alpha)$ space for $\eta_{\mathrm{edge}}=10$ (c) or $\eta_{\mathrm{edge}}=2$ (d).}
    \label{FS3}
\end{figure}

\newpage

\section{The convective threshold in two-dimensional systems}

In the main text, we analyzed a one-dimensional system, which is a reasonable approximation for anisotropic systems, such as striped patterns with a well-defined direction, or systems with planar structures, such as a flat interface separating bare soil from a vegetated area. Nevertheless, in more general scenarios, pattern formation in water-limited ecosystems is studied using two-dimensional models. Although the original non-local interaction-redistribution and reaction-diffusion water-biomass models are straightforward to generalize to two dimensions, generalizing the model reduction to two-dimensional scenarios is not as direct and requires making additional assumptions on the model parameters \cite{pinto2023topological}. Therefore, to show that our results also hold in higher dimensions, we introduce a phenomenological reduced equation for two-dimensional space that extends the one-dimensional limit we studied in the main text. 

To derive this phenomenological reduced equation, we note that the dominant contribution from anisotropy-induced nonreciprocal interactions in one dimension is $\alpha b \partial_x b(x,t)$. We then modified the two-dimensional reduced equation, originally derived for the isotropic case \cite{pintoramos2024arxiv}, by incorporating this term. The resulting two-dimensional model with phenomenological anisotropy-induced nonreciprocal interactions is
\begin{eqnarray}
    \frac{\partial b}{\partial t }= -\eta b +\kappa b^2-b^3 + (d-\Gamma b)\nabla^2b - \frac{1}{8}b\nabla^4b - \alpha b\frac{\partial b}{\partial x}, \label{Eq:2D}
\end{eqnarray}
where the parameters have the same meaning as in the one-dimensional equation, and the differential operators are given by $\nabla^2=\frac{\partial^2}{\partial x^2}+ \frac{\partial^2}{\partial y^2}$, $\nabla^4=(\nabla^2)^2$. The last term in Eq.\,\eqref{Eq:2D} accounts for the nonreciprocal interactions in the $x$ direction with intensity $\alpha$, and the $1/8$ factor defines an arbitrary spatial scale that has no influence on the dynamics \cite{Tlidi2008}.

Using Eq.\,\eqref{Eq:2D}, we repeated the analyses reported in the main text for the one-dimensional case. We varied the $\eta$ parameter quasi-statically for different combinations of the $(\Gamma, \alpha)$ parameters keeping $\kappa=0.6$, $d=0.02$ constant. We used a square simulation domain with $N=128$ points in each direction, $dx=2/3$, and Dirichlet boundary conditions. The results, shown in Fig.\,\ref{FS3}, indicate that the phenomenon of homogeneous covers being more resilient than patterned ones in the presence of nonreciprocities persists in the phenomenological two-dimensional model. Two bifurcation diagrams are exemplified in panels a) and b) for the homogeneous cover and patterned cover cases, respectively. In the absence of nonreciprocities, patterns can indeed resist high mortality rates; however, once nonreciprocities are included, the patterns collapse to bare soil for values of the mortality even lower than corresponding homogeneous covers. As in the main text, the pattern-formation is mainly controlled by the $\Gamma$ parameter, a proxy for the scale-dependent feedback mechanisms intensity. Sweeping the $(\Gamma, \alpha)$ parameter space, the negative effect of $\alpha$---decreasing always the mortality threshold before tipping---is unveiled, as seen in panel c). Taking two representative values of $\Gamma$, $\Gamma=0$ for homogeneous covers where Turing instabilities do not occur, and $\Gamma=0.5$ for patterned covers resulting from a Turing instability, one sees that there exist a critical nonreciprocity $\alpha_c$ after which patterns are less resilient than homogeneous covers, as seen in panel d). This provides an example in which our one-dimensional theory still applies to two-dimensional models. Nevertheless, it is still important to develop a formal, rigorous derivation of a reduced equation considering nonreciprocities in two-dimensional systems, and test the quantitative and qualitative features that higher dimensionality brings to our theory in future works.
\begin{figure}[]
    \centering
    \includegraphics[width=0.7\linewidth]{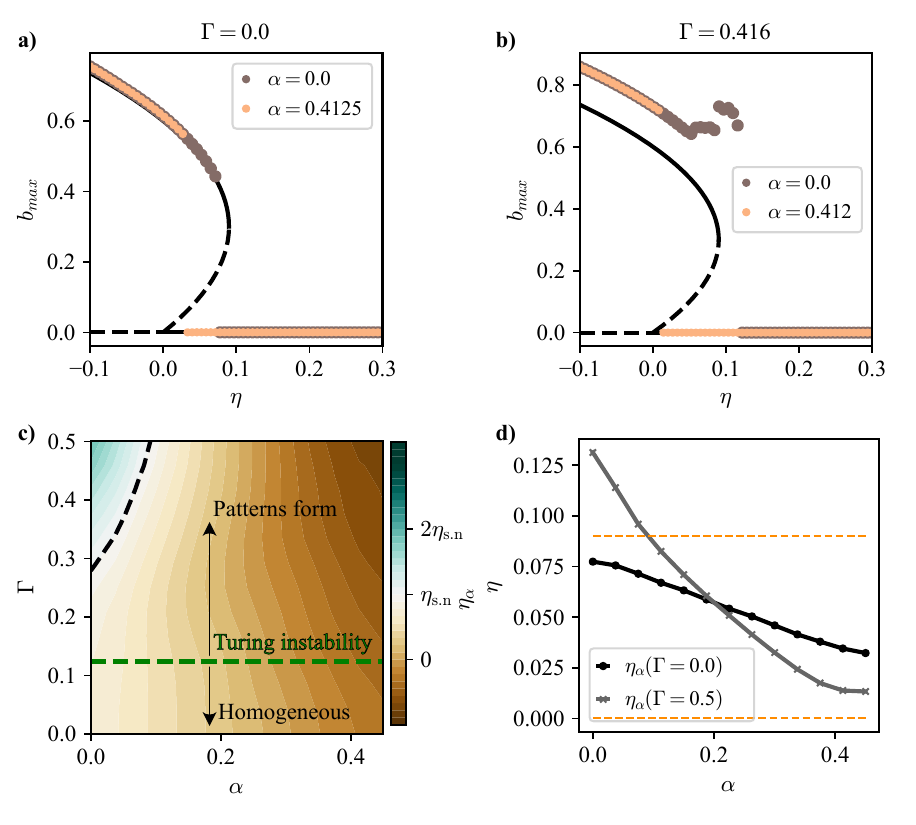}
        \caption[Convective instabilities triggered by nonreciprocities in a two-dimensional model]{Convective instabilities triggered by nonreciprocities in a two-dimensional model. a) bifurcation diagram for a homogeneous cover and different values of the nonreciprocity, $\alpha$. b) bifurcation diagram for a patterned cover and different values of $\alpha$. c) the tipping point induced by convective instabilities, $\eta_{\alpha}$, for different $(\Gamma, \alpha)$ values; the black dotted line indicates when $\eta_{\alpha}=\eta_\text{s.n}$ and the green dotted line shows the Turing instability threshold. d) tipping point value as a function of $\alpha$ for two illustrative $\Gamma$ values; $\Gamma=0$ for homogeneous covers, and $\Gamma=0.5$ for patterned covers.}
    \label{FS3}
\end{figure}

\newpage

\section{The order of magnitude of $\alpha_c$ in real ecosystems}

To test the plausibility of our theory as a mechanism of reduced resilience in real drylands, we estimated $\alpha_c$, the value at which anisotropy-induced nonreciprocities make patterned ecosystems less resilient than quasi-homogeneous ones, in a real ecosystem. For this, we considered the parameterization reported in Ref.\,\cite{meron2015ModelNew}, where the authors use the model Eq. \eqref{meron_model} in an isotropic environment ($V=0$) and for $D_B(B)=D_B$ to study potential regime shifts in fairy circles in Australia. Using that parameterization, we run our numerical protocol to obtain the nonlinear convective instability threshold in Eq.\,\eqref{meron_nondim} (note that it is non-dimensional) by quasi-statically varying the mortality parameter for different values of a hypothetical water advection velocity $V$. Then, the value of $V$ at which a patterned ecosystem becomes less resilient than a homogeneous one sets the critical anisotropy-induced nonreciprocity level. \\

\noindent \textbf{Model parameterization \cite{meron2015ModelNew}}. The parameters are: $E=7 \frac{\text{m}^2}{\text{year}}$, $K=0.4 \frac{\text{kg}}{\text{year}}$, $M=10.5 \; \text{year}^{-1}$, $L=15\; \text{year}^{-1}$, $R=0.9 \frac{\text{m}^2}{\text{kg}\cdot\text{year}}$, and $G=12 \frac{\text{m}^2}{\text{kg}\cdot\text{year}}$. In addition, the authors report precipitation rates rangin from $[50-150] \frac{\text{mm}}{\text{year}}$, which, using the upper limit and translating to density becomes $P=150  \frac{\text{kg}}{\text{m}^2\cdot\text{year}}$. The authors stimated the diffusion coefficient of water, $D_W$ to be in the range $[0.1-100] \frac{\text{m}^2}{\text{year}}$, and based on reference \cite{cain1997clonal} they used a value of $D_B=1.2 \frac{\text{m}^2}{\text{year}}$; however, the same reference gives various values for different plants. Based on that, we let $D_B$ to vary in the interval $[0.01- 1] \frac{\text{m}^2}{\text{year}}$. With these values, the nondimensional parameters of Eq. \eqref{meron_nondim} become: $\delta=2.8$, $\gamma=0.32$, $p=0.6$, $d_0=[0.01- 1] / D_W$, $d_1=0.4D_1/D_W$ (where, due to the lack of an estimate, we set $D_1=0.001$), $s=V/\sqrt{15D_W}$, and $m=0.7$. \\

\noindent \textbf{Computation of the nonlinear convective instability threshold.} To perform an analysis similar to the one presented in Fig. \ref{FS1}, we let the nondimensional mortality parameter, $m$ to vary between $[0.5-1.5]$, and sweep the simulation protocol over several values of $V[\frac{\text{m}}{\text{year}}]$ and the values of $D_B$ corresponding to different species. Note that the simulation is nondimensionalized, so we choose $D_W=1$ for the simulations, but the results can be extrapolated by running the conversion to dimensional units for any value of $D_W$. Plant species with low $D_B$ are more prone to develop spatial patterns than those with high $D_B$, so this parameter is equivalent to $\Gamma$ in the reduced equation. \\

\begin{figure}[]
    \centering
    \includegraphics[width=0.8\linewidth]{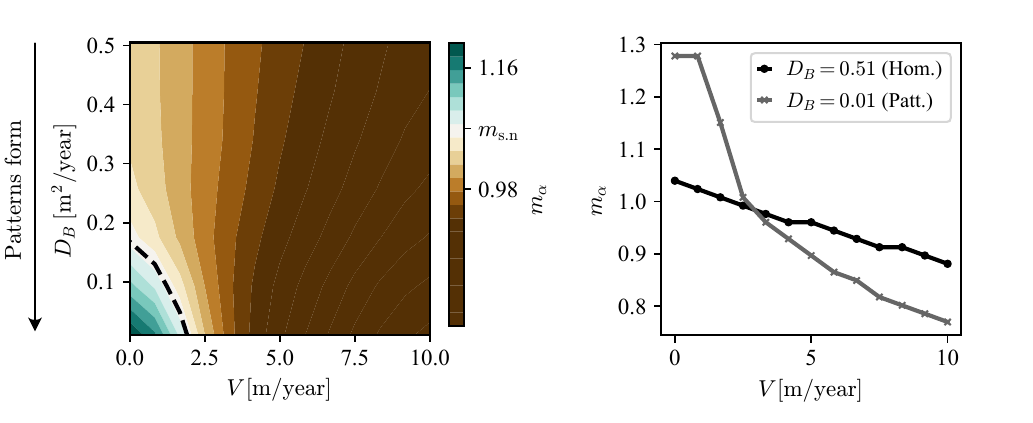}
        \caption[Simulation results for realistic parameter values]{Simulation results for realistic parameter values. The simulation protocol, increasing the nondimensional mortaity quasi-statically, was performed for various $(D_B, V)$ parameter combinations, and the units reconstructed assuming $D_W=1\;\text{m}^2/\text{year}$. The left panel shows the $m_{conv}$ threshold for different $(D_B, V)$ values compared to the non-spatial tipping point $m_c$. The right panel shows the value of $m_{conv}$ as a function of $V$ for two representative values of $D_B$ for which homogeneous or patterned covers were observed.}
    \label{FS4}
\end{figure}

\noindent \textbf{Results.} For the parameter values we considered, introducing water advection always reduces ecosystem robustness (Fig.\,\ref{FS4}). Moreover, there exists a value $V_c\approx 3 \frac{\text{m}}{\text{year}}$ after which patterned covers are less resilient than corresponding homogeneous ones. This estimated value is much lower than those reported in studies on anisotropic regions. For example, Klaumeier \cite{klausmeier1999regular} reports a water advection values of $V=365 \frac{\text{m}}{\text{year}}$. It is important to note, however, that this estimate is obtained in nondimensional units, which are reconstructed by assuming $D_W= 1 \frac{\text{m}^2}{\text{year}}$. More generally, the critical water advection velocity and plant dispersal rates can be rescaled using $V_c = V_c(D_W=1) \sqrt{D_W}\approx 3 \sqrt{D_W}$ and $D_B= D_B(D_W=1) D_W$. Using these expressions, the same results in Fig.\,\ref{FS4} would hold for water diffusion $D_W=100 \frac{\text{m}^2}{\text{year}}$, with $V_c= 30 \frac{\text{m}}{\text{year}}$ and plant dispersal rates between $[1-50] \frac{\text{m}^2}{\text{year}}$. The small value of isotropy-breaking parameters at which the model bifurcation diagram suggests that spatial patterning may reduced ecosystem resistance, supports the idea that non-reciprocities in plant interactions induced by weak anisotropies can be a source of reduced resilience in real ecosystems.

\end{document}